\documentclass[twocolumn]{aastex61}

\newcommand{\fodd}{$f_\mathrm{odd,Ba}$}
\newcommand{\foddr}{$f^r_{\mathrm{odd,Ba}}$}

\newcommand{\feh}{\mathrm{[Fe/H]}}

\newcommand{\cs}{CS\,29491-069}
\newcommand{\hen}{HE\,1219-0312}
\newcommand{\het}{HE\,2252-4225}
\newcommand{\hes}{HE\,2327-5642}

\received{July 1, 2017}
\revised{}
\accepted{}
\submitjournal{ApJ}

\shorttitle{The odd-isotope fractions of Barium in r-II stars}
\shortauthors{Cui et al.}
\begin{document}

\title{The odd-isotope fractions of Barium in the strongly r-process enhanced (r-II) stars\footnote{Based on observations
carried out at the European Southern Observatory, Paranal, Chile (Proposal number 170.D-0010 and  280.D-5011).}}

\correspondingauthor{Cui Wenyuan}
\email{wenyuancui@126.com, cuiwenyuan@hebtu.edu.cn}

\author[0000-0003-1359-9908]{Cui Wenyuan}
\affil{Department of Physics, Hebei Normal University, Shijiazhuang 050024, China}
\affil{School of Space Science and Physics, Shandong University at Weihai, Weihai 264209, China}
\author{Jiang Xiaohua}
\affiliation{Department of Physics, Hebei Normal University, Shijiazhuang 050024, China}
%\collaboration{(AAS Journals Data Scientists collaboration)}

\author{Shi Jianrong}
\affiliation{Key Lab of Astronomy, National Astronomical Observatories, Chinese Academy of Sciences, Beijing 100012, China}
\affiliation{School of Astronomy and Space Science, University of Chinese Academy of Sciences, Beijing 100049, China}
%\nocollaboration

\author{Zhao Gang}
\affiliation{Key Lab of Astronomy, National Astronomical Observatories, Chinese Academy of Sciences, Beijing 100012, China}
\affiliation{School of Astronomy and Space Science, University of Chinese Academy of Sciences, Beijing 100049, China}
%\collaboration{(LaTeX collaboration)}

\author{Zhang Bo}
\affiliation{Department of Physics, Hebei Normal University, Shijiazhuang 050024, China}
%\affiliation{American Astronomical Society \\
%2000 Florida Ave., NW, Suite 300 \\
%Washington, DC 20009-1231, USA}

%\author{Jeff Lewandowski}
%\affiliation{IOP Senior Publisher for the AAS Journals}
%\affiliation{IOP Publishing, Washington, DC 20005}



\begin{abstract}

We determined the \fodd\ values, $0.46\pm0.08$, $0.51\pm0.09$, 
$0.50\pm0.13$, $0.48\pm0.12$,  which correspond to the r-contribution 100\% 
for four r-II stars, \cs, \hen, \hes\, and \het, respectively.
Our results suggest that almost all of the 
heavy elements (in the range from Ba to Pb) in r-II stars have a common origin, that is, from
a single r-process (the main r-process). We found that the \fodd\ has a intrinsic nature, 
and should keep constant value of about 0.46 in the main
r-process yields, which is responsible for the heavy element enhancement of
r-II stars and of our Galaxy chemical enhancement. 
In addition, except the abundance ratio [Ba/Eu] the \fodd\ is also an important indicator,
which can be used to study the relative contributions of the r- and s-process during
the chemical evolution history of the Milky Way and
the enhancement mechanism in stars with peculiar abundance  of heavy 
elements.

\end{abstract}




\section{Introduction} \label{sec:intro}

%The detailed chemical abundances of stars provide the history informations of nucleosynthesis in our Galaxy. 
%Elements heavier than Iron ($\mathrm{Z} >30$) are mainly synthesized  through two neutron-capture processes, 
%i.e., rapid (r-) and slow (s-) process. 
The neutron-capture process produced almost all of the heavy elements ($\mathrm{Z} >30$)
in our universe, which was divided into the slow (s-) and rapid (r-) process, respectively
\citep{bur57}. The s-process produces about 82\% of solar barium and 6\% of solar europium, 
and the rest are produced by the r-process \citep{arl99}. Thus, barium is usually regarded as the representative element for 
the s-process, and europium for the r-process. 
%The s- and r-processes need very different conditions, 
%the former occurs in low-mass stars during their asymptotic giant branch phase with quietly helium-shell 
%burning \citep{bus99}, while the latter is associated with explosive conditions, such as type II 
%supernova \citep[a review see][]{thi02} and neutron star mergers \citep[see][and references therein]{wan14,gor15}.  
The s-process occurs in low- to intermediate-mass ($1-8\,M_\odot$) 
stars during their asymptotic giant branch phase with quietly helium-shell burning \citep{bus99}. 

Up to now, our knowledge about the r-processes is still poor.
The r-process is thought to be associated with explosive conditions such as type\,II 
supernova \citep[a review see][]{thi02} and neutron star mergers \citep[][and a review see Thielemann
et al. 2017]{wan14,gor15}, 
while it needs to be further identified.
Furthermore, the r-process yields are still not be well estimated with the current models for r-processes.
The current solar r-process abundances were obtained by subtract the s-process abundances
from the total solar abundances \citep{arl99}, which is also 
referred as r-residuals. \citet{bee05} defined stars with 
$\mathrm{[Eu/Fe]}>1.0$ and $\mathrm{[Ba/Eu]}<0$ as the r-II stars, and more than ten
r-II stars have been found to date \citep[see][and references therein]{cui13}, 
since \citet{sne94} reported the first one,
CS\,22892-052, with $\mathrm{[Eu/Fe]}=1.6$ and $\mathrm{[Ba/Eu]}=-0.7$.
Obviously, study of the r-II stars can greatly improving our knowledge on
the r-process especially early in the Milky Way. 
%\citet{sne94} found that the ``main r-process" star, CS\,22892-052, exhibits strong r-process signature 
%with $\mathrm{[Eu/Fe]}=1.6$ and $\mathrm{[Ba/Eu]}=-0.7$, which was later classed as the r-II stars, i.e., 
%with $\mathrm{[Eu/Fe]}>1.0$ and $\mathrm{[Ba/Eu]}<0$, by \citet{bee05}. Since 1994, more than ten 
%metal-poor stars have been found to belong to the r-II groups  \citep[see][and references therein]{cui13}. 

In fact, almost all of the r-II stars found up to now show well consistence between their heavier 
element abundances ($Z \geq 56$) and the scaled solar pure r-process abundances 
\citep[see][and references therein]{sne08,cow11}, while their light neutron-capture elements 
($37\leq Z\leq47$, i.e., from Rb to Ag) are more deficient than the solar r-process ones \citep{cow06}. 
This implies that different sites and different r-processes are responsible for the solar pure r-process abundances 
of heavier and lighter neutron-capture elements, respectively. 
%It also indicates that the pattern of the solar 
%pure r-process abundances for heavy elements is universal through large metallicity range, 
%in other words, it shows a robust nature on the abundance level. 
Thus, the main r-process (i.e., the classical r-process) was suggested to mainly explain the 
solar r-process abundances of heavier neutron-capture elements \citep{rya96,sne08},
and the ``lighter element primary process'' (LEPP) \citep{tra04} or Òweak r-processÓ \citep{ish05}
was suggested to mainly produce the lighter neutron-capture elements.
The good agreements between the heavier 
element abundances and the scaled solar pure r-process
indicate that the solar r-process pattern of heavier elements 
is robust and universal through large metallicity range in the Galaxy.

Different isotopic abundance mixtures of a neutron-capture element are produced by the r- and the s-processes.
For instance, Eu and Ba have two and five stable isotopes, i.e., $^{151}$Eu, $^{153}$Eu, and  
$^{134}$Ba, $^{135}$Ba, $^{136}$Ba, $^{137}$Ba, $^{138}$Ba, respectively. \citet{arl99} predicted 
that $f_{151}$ ($=N(^{151}\mathrm{Eu})/N(\mathrm{Eu})$) is about the same value 0.47 in the 
r-process component of the solar Eu abundance ($f^r_{151}$ hereafter) with
their ``stellar" and ``classical" models, and  about 0.54 and 0.59 in the s-process 
component of the solar Eu abundance ($f^s_{151}$ hereafter), respectively. 
For Ba, the ``stellar" model gave $f^{r/s}_\mathrm{odd,Ba}$\footnote{
$f^{r/s}_\mathrm{odd,Ba}= [N(^{135}\mathrm{Ba})+N(^{137}\mathrm{Ba})]/N(\mathrm{Ba})$,
where r represents \fodd\ in the r-process component of the solar Ba abundance, 
s represents \fodd\ in the s-process component of the solar Ba abundance.}
$\simeq 0.46$ and 0.11 in the r- and s-process components of the solar Ba abundance, 
while the ``classical" model gave $f^{r/s}_\mathrm{odd,Ba}$\,$\simeq 1.0$ and 0.09, respectively.   
Obviously, large difference arise from the \fodd\ in the r-components of the solar Ba abundance 
predicted by the  ``stellar" and ``classical" models, respectively.
Eu isotopic fractions, $f_{151}$, in several r-II stars have been reported 
by \citet{sne02} for CS\,22892-052 ($0.5\pm0.1$), HD\,115444 ($0.5\pm0.1$), BD\,+17$^\circ$3248 ($0.5\pm0.1$), 
and \citet{aok03} for CS\,31082-001 ($0.44\pm0.1$), CS\,22892-052 ($0.51\pm0.1$), 
HD\,115444 ($0.46\pm0.1$). 
%which is in agreement with the values of solar system 
%material such as 0.478 from the meteorolite \citep[e.g.,][]{lod03} and $0.50\pm0.07$ from 
%the solar photosphere \citep{law01}. In fact, Eu isotopic fractions, $f_{151}$, are 0.47 and 0.54 
%for the solar pure r- and s-process, respectively, predicted by the stellar model \citep{arl99}. 
%Combining the obtained stellar values of $f_{151}$ for some r-II stars and the theoretical 
%Considering the small interval between the $f_{151}$ values in the 
%r- and s-process component of the solar Eu abundance, about 0.07\,dex (``stellar model") or 
Considering the small interval between the values of $f^r_{151}$ and $f^s_{151}$, about 0.07\,dex (``stellar model") or 
0.12\,dex (``classical model"), and the relative large errors of $f_{151}$ measured from r-II stars,
the $f_{151}$ maybe only used to marginally distinguish the r- and s-process 
contributions to these stars \citep[also see Figure\,13 of][]{sne08}. 
Comparing to Eu, the large interval ($\sim0.35$\,dex) between the values of $f^r_{\mathrm{odd,Ba}}$ 
and $f^s_{\mathrm{odd,Ba}}$
may support to get reliable conclusions. The Ba 
isotope fraction was first reported by \citet{mag93} to study the r-process contribution to 
the Ba abundance in the metal-poor ($\feh\,=-2.4$) subgiant HD\,140283, their result is 
subsequently examined by a series of works \citep{mag95,col09,lam02,gal10,gal12,gal15}. 
However, contrary conclusions have been obtained with different \fodd\ values determined 
by the above works. The main reasons are the low barium abundance, $\mathrm{[Ba/Fe]}=-0.87$  
(also weak Ba lines) in HD\,140283. In addition, we note that the non-local thermodynamical equilibrium 
(NLTE hereafter) effects have not been considered during analyses of Ba isotope fractions in the above works, 
which may also lead to some uncertainties on this issue. 
Furthermore, \citet{mas99,sho06} have confirmed that the profile of the Ba\,II resonance 
line $\lambda4554$ (used to measure \fodd) suffers strong NLTE effects for metal-poor stars. \citet{mas06} 
have determined the Ba odd-isotope fractions for 25 disk stars, and found that the thick disk 
stars show larger \fodd\ values than these of the thin disk stars, and their \fodd\ values decrease 
with increasing metallicities.
Ba isotope fractions in r-II stars were first reported by \citet{men16}, who found that the \fodd\ 
values in CS\,31082-001 is consistent with the isotopic ratio in the solar pure r-process component. 
%In fact, CS\,31082-001 show higher abundances of the third r-process peak ($Z\ge72$) 
%than the scaled solar r-residuals compared to other r-II stars \citep{hil02}. 

In order to give 
a unambiguous verdict on the origin of heavy elements in r-II stars and giving critical constraint on
theoretical models for the r-process, on the isotopic level, it should also be important
to determine the isotopic fractions of other heavy elements especially 
Ba (surrogate of the s-process) except Eu for a large samples of r-II stars.
%except Eu the determination of the isotopic fractions for other heavy elements especially 
%Ba (surrogate of the s-process) and that in a large samples of r-II stars should also be important. 
In this paper, 
we report the Ba odd-isotope fractions for four stars with strongly r-process enhancement, 
CS\,29491-069, HE\,1219-0312 , 
HE\,2327-5642 and HE\,2252-4225, and this is also the first report on isotopic 
fractions of their heavy elements. The paper is structured as follows. After presenting 
the observational data in Sect. 2, we describe model atmospheres and stellar parameters
determination in Sect. 3, and the results are showed in 
Sect. 4. The origin of heavy element for r-II stars is analyzed in Sect. 5,
and the characters of \fodd\ in r-II stars are described in Sect. 6, while our conclusions are presented in Sect. 7.

\section{Observations and Data Reduction} \label{sec:obs}

For convenience, we list astrometry and photometry informations of the four r-II stars \cs, \hen, \hes\ and \het\ 
in Table~\ref{astrometry}. The photometry result was taken from \citet{bee07}.
All  the high resolution, high signal-to-noise spectra of these stars were observed with the VLT spectrograph 
UVES. \cs\ and \hen\ were observed with resolutions of $\sim 60\,000$ and $\sim 70\,000$ 
in 2004 and 2005 by \citet{hay09}, respectively. The observations of \cs\ and \hen\ were 
spanned $\sim 1.5$ and $\sim 14$ months, in which the total integration times are 1\,h 
and 15\,h for the spectra including the Ba\,II resonance line at 4554\,\AA, while 1\,h 
and 16\,h of exposures for the region with Ba\,II lines at 5853 and 6496\,\AA, respectively 
\citep[detailed see][in their Table 2]{hay09}. The $S/N$ of the final combined spectra 
are 70 and 50 around 4554\,\AA, 110 and 111 around 5853 and 6496\,\AA\ for \cs\ and \hen, respectively.

\begin{table*}%\small
\begin{center}
\caption{Astromety and photometry of the sample stars}
\label{astrometry}
\begin{tabular}{lcccc}
\hline
\hline
Quantity & $\cs\ $ & \hen\ & \hes\  & \het\   \\
\hline
RA (2000.0)        & 22:31:02.1  & 12:21:34.1& 23:30:37.2 & 22:54:58.6 \\
dec (2000.0)        & -32:38:36  & -03:28:40 & -56:26:14 & -42:09:19 \\ 
$V$ [mag]      &  $13.075\pm0.002$ & $15.940\pm0.007$ & $13.881\pm0.003$ &$14.878\pm0.003$  \\
$B-V$             &  $0.600\pm0.004$&  $0.641\pm0.027$ &  $0.709\pm0.005$ & $0.822\pm0.005$  \\
$V-R$      &  $0.421\pm0.003$ & $0.455\pm0.011$ & $0.456\pm0.004$ &$0.499\pm0.005$  \\
$V-I$         &  $0.900\pm0.004$ & $0.897\pm0.009$ & $0.933\pm0.005$ &$1.023\pm0.006$  \\
\hline
\end{tabular}
\end{center}
\end{table*}

High-quality spectra of \hes\ and \het\ were obtained during May-November 2005 
with $R\sim60\,000$ and $R\sim50\,000$, respectively. For \hes, the total integration time 
are 10\,h and 4\,h for the spectra at $4554$\,\AA\ and $5853$\AA, $6496$\,\AA, respectively. 
The $S/N$ of the resulting spectra are larger than 50 around 4554\,\AA, and 100 around 5853, 6496\,\AA, 
respectively \citep[detailed see][in their Table 2]{mas10}. For \het, the total integration time are 10\,h 
and 9\,h for the spectra at $4554$\,\AA\ and $5853$\,\AA, $6496$\,\AA, respectively. 
The $S/N$ of the resulting spectra are larger than 70 around 4554\,\AA, and 60 around 5853 
and 6496\,\AA, respectively \citep[detailed see][in their Table 2]{mas14}. The raw 
data of the above four r-II stars were downloaded from the European Southern 
Observatory (ESO) archive\footnote{http://archive.eso.org/cms.html}. 

The spectra were reduced with an IDL software package, which was designed originally for the 
FOCES spectrograph \citep{pfe98}, and has been modified to work for UVES spectrograph.
Cosmic rays and bad pixels were
removed by careful comparisons of different exposures from the same objects. The instrumental 
response and background scatter light were also considered during the data reducing.
It is worth noting that we have not found the proper lamp exposures with the same settings for the red region 
taken by the CD\#3 equipped on the red arm of the UVES spectrograph for almost all of the sample stars except \het, 
while we have found them for the blue region for \cs, \hen, and \het\ except \hes. 
Thus, the pipeline-reduced spectra have been adopted for the red 
spectra region ($>$\,500\,nm) for \cs, \hen, and \hes, which also was adopted for the blue region ($< 500$\,nm) for \hes.

\section{Model Atmospheres and Stellar Parameters} \label{sec:par}

We redetermined the stellar parameters of the four r-II sample stars in this work in order to 
reduce the uncertainties of the final results, although they can be found in literatures \citep{hay09,mas10,mas14}.
%Although the latest stellar parameters of the four r-II sample stars can be found in literatures \citep{hay09,mas10,mas14}, 
%which for \cs\ and 
%\hen\ was determined by \citet{hay09} with model atmospheres MARCS adopted and LTE 
%hypothesis considered, and which for \hes\ and \het\ was determined by \citet{mas10,mas14} with model atmospheres 
%MAFAGS adopted and NLTE effects considered.
%we redetermined them in this work in order to reduce the uncertainties of the final results.   
%\subsection{Model Atmospheres} \label{sec:mod}
The opacity sampling (OS) model atmosphere MAFAGS-OS9 was adopted in the 
following analysis processes of the stellar parameters, the Ba abundance and the Ba isotopic fraction. 
This model was developed by \citet{gru04} and updated by \citet{gru09}, which is based on  
one-dimensional plane-parallel (LTE) model. In the new version, the new iron atomic data 
computed by \citet{kur09} was incorporated.  

\begin{table*}\small
\begin{center}
\caption{Comparison of stellar parameters with other studies}
\label{stellardata}
\begin{tabular}{lccccl}
\hline
\hline
Name & $T_\mathrm{eff}$ & $\log g$ & [Fe/H]  & $V_{mic}$ & Referencens\tablenotemark{a}  \\
 &(K) & (dex)& (km\,s$^{-1}$) & &\\
\hline
\cs        & $5200$  & $2.80$ & $-2.58$ & $1.8$ & This work \\
             & $5300$  & $2.80$ & $-2.51$ & $-1.6$ & HAY09 \\
%             & 5103  & 2.50 & $-2.76$ & $-1.5$ & BAR05 \\             
\hen      &  5060 & 2.30 & -2.94 &1.7 & This work  \\
             &  5060 & 2.30 & -2.96 &1.6 & HAY09  \\
%             &  5140 & 2.40 & -2.80 &1.5 & BAR05  \\
\hes      &  5000 & 2.34 & -2.87 &1.6 & This work  \\
             &  5050 & 2.34 & -2.78 &1.8 & MAS10  \\
\het       &  4750 & 1.65 & -2.67 &1.7 & This work  \\
             &  4710 & 1.65 & -2.63 &1.7 & MAS14  \\          
\hline
\end{tabular}
\end{center}
\tablenotetext{a}{ HAY09: \citet{hay09}, %BAR05: \citet{bar05}, 
MAS10: \citet{mas10}, MAS14: \citet{mas14}.}
\end{table*}

%\subsection{Stellar Parameters} \label{sec:stel}
We derived the stellar parameters of the four r-II stars via the spectroscopic approach, namely, 
the effective temperature $T_\mathrm{eff}$ was determined by requiring the equilibrium 
between the excitation and iron abundance from Fe\,I lines, and the surface gravity 
$\mathrm{log}\,g$ was derived from the ionization equilibrium of Fe\,I and Fe\,II. 
The microturbulence velocity $V_\mathrm{mic}$ was estimated by requiring [Fe/H] derived from 
the Fe\,I lines to be independent of their equivalent widths, and the equivalent widths are listed 
in Table~\ref{tab:linedata}. An iterative procedure was adopted to measure stellar parameters, 
where the initial parameters were taken from the latest literatures, e.g., \citet{hay09}
for \cs\ and \hen, \citet{mas10} for \hes\ and \citet{mas14}  for \het. 
%The adopted initial parameters
%for \cs\ and \hen\ were determined by \citet{hay09},  and those for \hes\ and \het\ were derived
%by \citet{mas10} and \citet{mas14}, respectively.
The initial effective temperatures were all derived by fitting the Balmer-line profiles of both 
H$\alpha$ and H$\beta$ using the manually rectified spectra, while the initial
surface gravities were determined through the ionization equilibrium of Fe\,I and
Fe\,II. There are 21 Fe\,I and 5 Fe\,II lines included in our analysis, which were listed in Table~\ref{tab:linedata} with their 
atomic parameters and the references of the adopted $gf$-values. 
The van der Waals broadening of the iron lines is accounted for using the most accurate data available 
as provided by \citet{ans95,bar97,bar98,bar00,bar05}.
The final stellar parameters 
are given in Table~\ref{stellardata}. During this process, the NLTE line formation for Fe\,I and 
Fe\,II have been considered, where the iron model atom was adopted from \citet{mas11}.
Based on multiple iterative processes, the typical
uncertainties of $T_\mathrm{eff}$, $\mathrm{log}\,g$, [Fe/H] and $V_\mathrm{mic}$ are
estimated to be $\pm80$\,K, $\pm0.20$\,dex, $\pm0.08$\,dex and $\pm0.1$\,km\,s$^{-1}$, respectively.%,
%except $\mathrm{log}\,g$ for \het\ and \hes, which is $\pm0.15$\,dex.

For comparing, the stellar parameters for the sample stars from the literature are also presented in 
Table~\ref{stellardata}. There are good consistence between our  
stellar parameters and those from literatures, except the effective temperature 
$T_\mathrm{eff}$ for \cs. The newly determined $T_\mathrm{eff}$ value is 
5200\,K, which is 100\,K lower than that derived by \citet{hay09}. 
%However, we also noted that 
%the uncertainties of $T_\mathrm{eff}$ for \cs\ were estimated to be $\pm80$\,K and $\pm100$\,K 
%by this work and \citet{hay09}, respectively. This may indicate that
%the value of $T_\mathrm{eff}$ for \cs\ determined in this work is
%consistent with that derived by \citet{hay09}. 
However, considering the uncertainties of $T_\mathrm{eff}$ for \cs\, 
$\pm80$\,K estimated by this work, and $\pm100$\,K by \citet{hay09},
our $T_\mathrm{eff}$ is still consistent with
that from \citet{hay09} in some degree.  
%the uncertainties of $T_\mathrm{eff}$ for \cs\ were estimated to be $\pm80$\,K and $\pm100$\,K 
%by this work and \citet{hay09}, respectively. This may indicate that
%the value of $T_\mathrm{eff}$ for \cs\ determined in this work is
%consistent with that derived by \citet{hay09}. 
The good consistence between the stellar
parameters determined by this work and those from literatures \citep{hay09,mas10,mas14} may be mainly due 
to the considering of NLTE effects for Fe\,I and Fe\,II lines in their determining processes.

\section{Results\label{sec:result}}

\subsection{Methods}

Ba has five stable isotopes, i.e., $^{134}$Ba, $^{135}$Ba, $^{136}$Ba, $^{137}$Ba, $^{138}$Ba, 
where $^{134}$Ba and $^{136}$Ba, are s-only nucleus due to the shielding by $^{134}$Xe 
and $^{136}$Xe on the r-process path, the others are produced by both the r- and s-process 
nucleosynthesis. Therefore, the Ba odd-isotope fractions can help us to 
estimate the r- and s-process contributions to the heavy elements in a star, e.g. larger \fodd\, 
values corresponding to larger r-process contributions. 

In order to study the nature of the neutron-capture processes in r-II stars, following
the approach adopted by \citet{mas06} and \citet{men16}
we determined the fractions of the odd Ba isotopes \fodd\ for \cs, \hen\, \het\ and \hes.
%We follow the approach adopted by \citet{mas06,men16} to deriving the \fodd\ values.
This method is firstly to derive the stellar total Ba abundances from lines insensitive to 
HFS or isotope shifts, and two subordinate lines of Ba\,II at 5853 and 6496\,{\AA} 
were adopted here. The average Ba abundances derived from the above two lines
were adopted as the final total Ba results.
Subsequently, the \fodd\ values was determined through fitting the line profile
of the Ba\,II resonance line at 4554\,\AA. This line is sensitive to the Ba odd-isotope
fractions. During this process, the stellar total Ba abundance was fixed, and 
the \fodd\ values changed freely until the best fitting result obtained, and simultaneously
the final \fodd\ value was obtained. In this work, the adopted Ba atomic model is same as that
adopted by \citet{mas99} and \citet{mas06}.
The IDL/Fortran SIU software package of \citet{ree91} 
%The \software{SIU(Reetz 1991)} of IDL/Fortran package
was used to compute the synthetic line profiles. 

\startlongtable
\begin{deluxetable*}{lcccccccc}
\tablecaption{Line data for Fe\,I and Fe\,II lines used to determine the stellar parameters, and 
equivalent widths of neutral iron lines for sample stars (EW$_1$: \cs, EW$_2$: \hen, EW$_3$: \hes, EW$_4$: \het)\tablenotemark{a}. \label{tab:linedata}}
\tablehead{
\colhead{$\lambda\,(\AA)$} &\colhead{$\chi_{ex}$\,(eV) } & \colhead{$\mathrm{log}\,gf$} &\colhead{$\mathrm{log}\,C_6$} &
 \colhead{EW$_1$} &  \colhead{EW$_2$} & \colhead{EW$_3$} & \colhead{EW$_4$} & \colhead{Reference} %\\
%\colhead{} & \colhead{cost} & \colhead{charges\tablenotemark{b}}\\
%\colhead{} & \colhead{(\$)} & \colhead{(\$/page)}
}
%\colnumbers
\startdata
Fe\,I \\
4427.310 & 0.05 & -2.92 & -31.86 & 80.3 & 67.7 & 83.0 & 105.1 & OB91 \\                 
4920.505 & 2.83 &  0.07 & -30.51 & 70.8 & 51.4 & 59.0 &  84.4 & OB91 \\                 
4994.129 & 0.91 & -2.96 & -31.71 & 33.2 & 24.9 & 29.0 &  60.8 & BA91 \\                 
5166.282 & 0.00 & -4.20 & -31.93 & 35.6 & 24.1 & 30.3 &  65.8 & BL79 \\                 
5198.717 & 2.22 & -2.14 & -31.32 & 13.8 & $\cdots$  & 11.6 & $\cdots$  & BL82a \\       
5216.274 & 1.61 & -2.15 & -31.52 & 42.7 & 29.8 & 33.5 & $\cdots$  & FU88 \\             
5232.940 & 2.94 & -0.06 & -30.54 & 59.4 & 48.5 & 54.1 & $\cdots$  & OB91 \\             
5247.056 & 0.09 & -4.95 & -31.92 & $\cdots$ & $\cdots$ & $\cdots$ &  14.5  & BL79 \\    
5281.791 & 3.04 & -0.83 & -30.53 & $\cdots$  & 17.1 &  $\cdots$  & $\cdots$ & OB91 \\   
5324.180 & 3.21 & -0.10 & -30.42 & 43.3 & 27.7 & 33.5 &  55.1 & BA91 \\                 
5367.470 & 4.41 &  0.44 & -30.20 & 14.2 & $\cdots$ & $\cdots$ & $\cdots$ & OB91 \\      
5383.369 & 4.31 &  0.64 & -30.37 & 22.9 & 19.2 & $\cdots$ &  $\cdots$   & OB91 \\       
5393.173 & 3.24 & -0.72 & -30.42 & 20.2 & 12.6 &  $\cdots$ & $\cdots$ \ & BA91 \\       
5434.530 & 1.01 & -2.12 & -31.74 & 77.9 & 64.8 & 77.8 &  105.5 & FU88 \\                
5586.760 & 3.37 & -0.10 & -30.38 & 34.7 & 22.9 & 24.6 &  48.0  & BA91 \\                
6065.490 & 2.61 & -1.53 & -31.41 & 20.4 & $\cdots$ & 11.7 &  26.5 & BL82b \\            
6213.430 & 2.22 & -2.48 & -31.58 &  $\cdots$  & $\cdots$  &  $\cdots$  &  12.8 & OB91 \\
6252.560 & 2.40 & -1.69 & -31.52 & 23.3 & 12.2 & 18.4 &  35.7 & BL82a \\                
6393.612 & 2.43 & -1.43 & -31.53 & 26.8 & $\cdots$ & 20.4 &  44.0 & BL91 \\             
6411.649 & 3.65 & -0.60 & -30.38 & $\cdots$ & $\cdots$ & $\cdots$ &  12.0 & BL91 \\     
6421.351 & 2.28 & -2.03 & -31.80 & 18.0 & $\cdots$  & $\cdots$ &  26.5 & BL82a \\       
\hline
Fe\,II \\
4491.404 & 2.84 & -2.76 & -32.02  &$\cdots$&$\cdots$&$\cdots$&$\cdots$& RA98 \\
4508.288 & 2.84 & -2.34 & -32.00  &$\cdots$&$\cdots$&$\cdots$&$\cdots$& RA98 \\
4923.927 & 2.88 & -1.42 & -32.03  &$\cdots$&$\cdots$&$\cdots$&$\cdots$& MO83 \\
5018.440 & 2.88 & -1.23 & -32.04  &$\cdots$&$\cdots$&$\cdots$&$\cdots$& MO83 \\
5197.577 & 3.22 & -2.34 & -32.02  &$\cdots$&$\cdots$&$\cdots$&$\cdots$& RA98 \\
\enddata
\tablenotetext{a}{Strongly blended lines and lines with poor $S/N$ were rejected when the stellar parameters were determined for a given star.}
%\tablenotetext{b}{Accounts for the change from page charges to digital quanta in April, 2011}
\tablecomments{References for the $gf$-values are BA91: \citet{bar91}, BL79: \citet{bla79}, 
BL82a: \citet{bla82a}, BL82b: \citet{bla82b}, FU88: \citet{fuh88}, MO83: \citet{moi83},
OB91: \citet{obr91}, RA98: \citet{raa98}.}
\end{deluxetable*}

\begin{table*}[!htbp]
\begin{center}
\caption{Line data, equivalent widths (EW, unit: m\AA), and NLTE barium abundances of individual Ba\,II lines for the sample stars.}
\label{ba.abundance}
\begin{tabular}{lcccccccccc}
\hline
\hline
$\lambda$\,(\AA) & $\chi_{ex}$\,(eV) & $\log gf$ &EW & [Ba/Fe] &EW& [Ba/Fe] &EW& [Ba/Fe] &EW& [Ba/Fe]   \\
\hline
\multicolumn{3}{c}{} &\multicolumn{2}{c}{\cs}&\multicolumn{2}{c}{\hen}&\multicolumn{2}{c}{\hes}&\multicolumn{2}{c}{\het}\\
\hline
5853.668   & 0.604 & -1.000 &20.7& 0.27 & 37.9 &0.74&25.1&0.37&44.9&0.17\\
6496.897   & 0.604 & -0.377 &58.1& 0.25 & 75.8&0.69&62.1 &0.31&91.4&0.14\\
4554.029  & 0.000 & 0.170  &124.1 &--&136.9&--&127.2&--&151.8&--\\
\hline
%\vspace{-11mm}
\end{tabular}
\end{center}
%\tablenotetext{a}{ The unit of EW is m\AA.}
\end{table*}

\begin{table*}%\small
\begin{center}
\caption{NLTE Ba abundances and Ba odd-isotope fractions for the sample stars, respectively.}
\label{nltefodd}
\begin{tabular}{lcccccc}
\hline
\hline
Name &[Ba/Fe]\tablenotemark{L} & [Ba/Fe]  & $\sigma$(total)&  \fodd & $\Delta$ \\
\hline
\cs\     &  0.24 & $0.26$&$0.13$ & $0.46$ & 0.08\\
\hen    &0.65  & $0.72$ &$0.10$ & $0.51$ & 0.09\\
\hes    & 0.31 & $0.34$&$0.05$ & $0.50$ & 0.13\\
\het     & 0.29 & $0.16$ &$0.08$ & $0.48$ & 0.12\\
\hline
\end{tabular}
\tablenotetext{L}{ LTE abundance}
\end{center}
\end{table*}

The reduced  $\chi^2$, $\chi_\mathrm{r}^2$\,\footnote{The reduced $\chi^2$ is defined as
%\begin{center}
\begin{eqnarray*}
\chi_r^2=\frac{1}{\nu-1}\sum_{i=1}\frac{(O_i-S_i)^2}{\sigma_i^2},
\end{eqnarray*}
where $O_i$ is the observed continuum-normalised flux, $S_i$ is the 
synthetic spectral points, $\nu$ is the number of degrees of freedom in 
the fit, and $\sigma_i$ is the 
standard deviation of the data points defining the continuum of the observed spectrum \citep{smi98}. 
$\sigma$ is defined as $\sigma=(S/N)^{-1}$, where $S/N$ is measured in roughly 1\,\AA\ interval on 
either side of the spectral line referred.} 
was computed to find the best fit to the observed spectra from a set of 
synthetic ones. Same as that in \citet{men16}, there are
three free parameters: the wavelength shift, a continuum level
shift to match the synthetic continuum, and  macroturbulence when determining
the \fodd\ values, which are required to make the comparison between the synthetic and 
observed spectra. A careful renormalisation for the observed spectra was done first
to fix the continuum level over a window of each Ba\,II resonance line at 4554\,\AA.
In our fitting, the line profiles of $\lambda$\,4554\,\AA\ include 22, 25, 23, and 27\, pixels 
for \cs, \hen, \hes\ and \het, respectively. Thus the total degrees of freedom $\nu$
are adopted as 22, 25, 23, and 27 minus two fitting parameters for the sample stars,
respectively. The minimum value is expected for $\chi_\mathrm{r}^2$ to get
the best fit results. \citet{smi98} had pointed out that any non-Gaussian extended 
wings of the instrumental profile are weak, which could be ignored during the 
line-profile fitting. The instrumental broadening was thus determined from a Th-Ar
lamp spectrum by a Gaussian fit, which was observed with the same instrumentation setup 
when the object exposures. The rotation broadening is also involved in our 
synthetic profiles. \citet{smi98} had shown that $v\,\mathrm{sin}\,i$ is less than 3\,km\,s$^{-1}$ 
for old stars, where $v$ is the surface equatorial rotational velocity. In this work,
we adopted $v\,\mathrm{sin}\,i=1.5$\,km\,s$^{-1}$ as the projected rotational velocity
for our sample stars.

%% The "ht!" tells LaTeX to put the figure "here" first, at the "top" next
%% and to override the normal way of calculating a float position
%\begin{figure}[ht!]
%\plotone{cost.eps}
%\caption{The subscription and author publication costs from 1991 to 2013.
%The data comes from Table \ref{tab:table}.\label{fig:general}}
%\end{figure}

\subsection{Stellar Total Ba Abundances \label{abun}}

For the sample r-II stars, Ba abundances were obtained through fitting the line
profiles of Ba\,II $\lambda5853$ and $\lambda6496$, respectively. The NLTE
effects were also considered. Table~\ref{ba.abundance} shows the line data of Ba\,II lines
$\lambda5853$, $\lambda6496$ and $\lambda4554$, respectively. The equivalent widths and
Ba abundance from individual line for our sample stars were also presented in Table~\ref{ba.abundance}. 
The final Ba abundances were provided in Table~\ref{nltefodd},
and the LTE Ba abundances were also included for comparing.

\begin{figure*}
\gridline{\fig{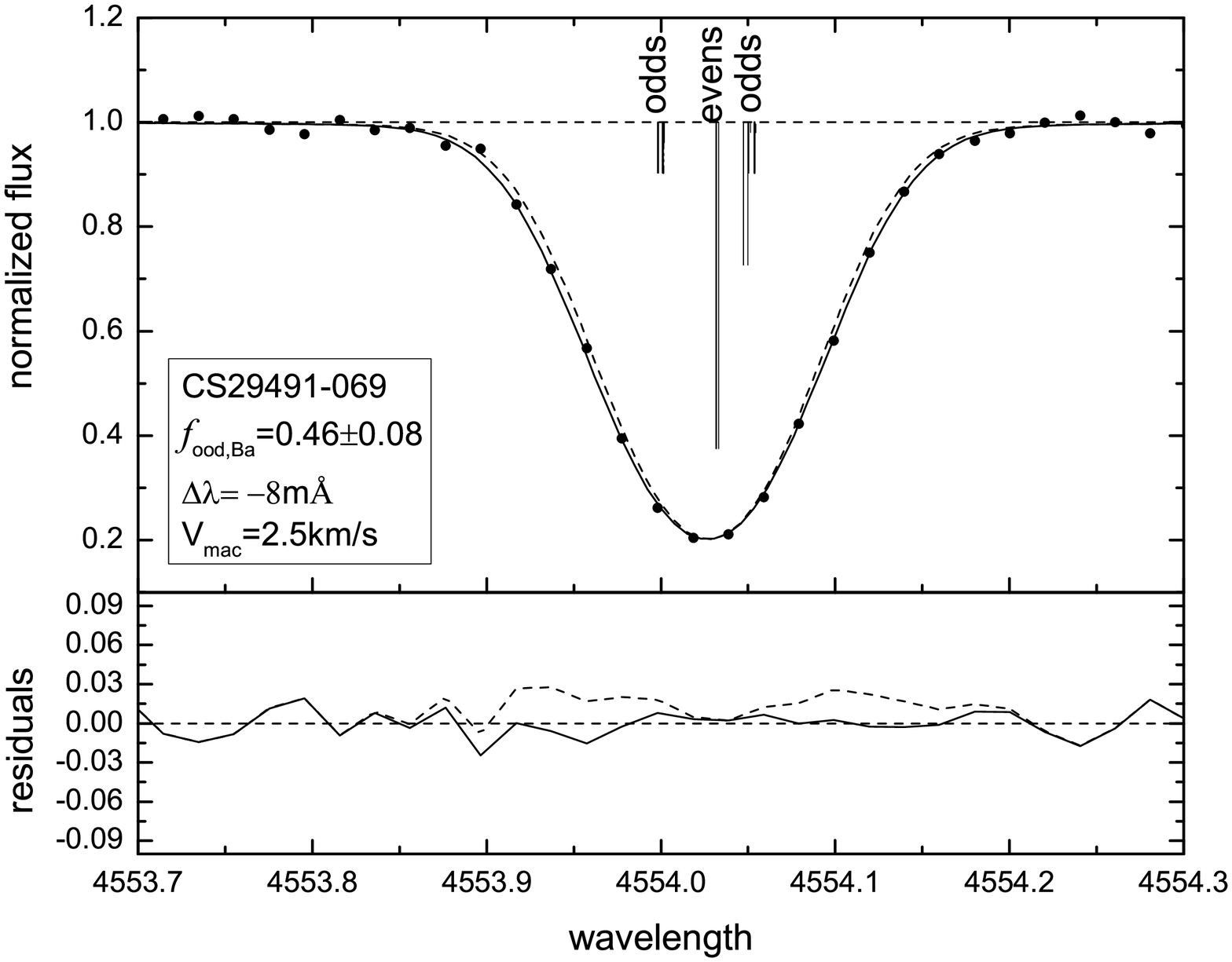}{0.4\textwidth}{}
 \hspace{-32mm}
          \fig{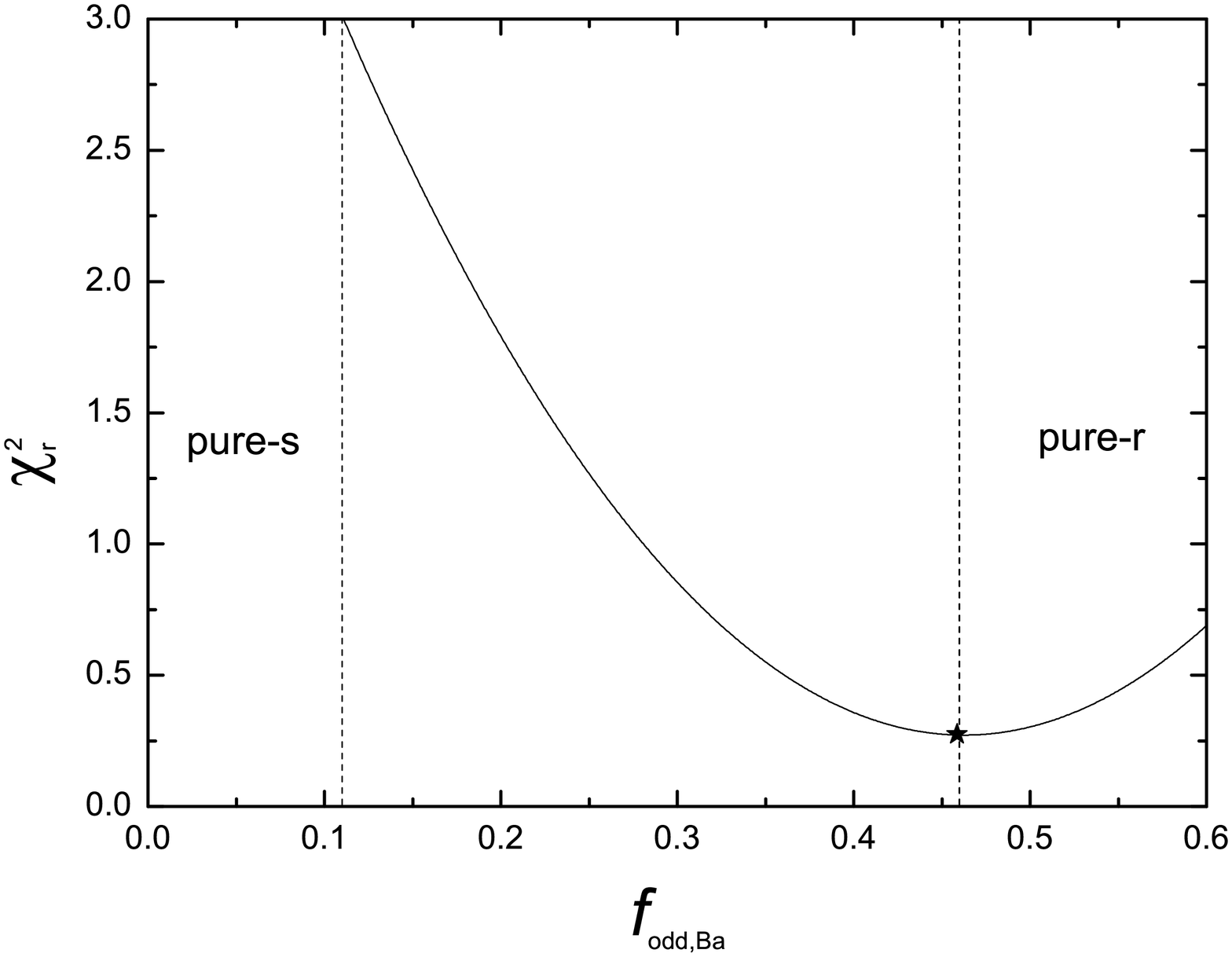}{0.4\textwidth}{}
          }
           \vspace{-15mm}
\gridline{\fig{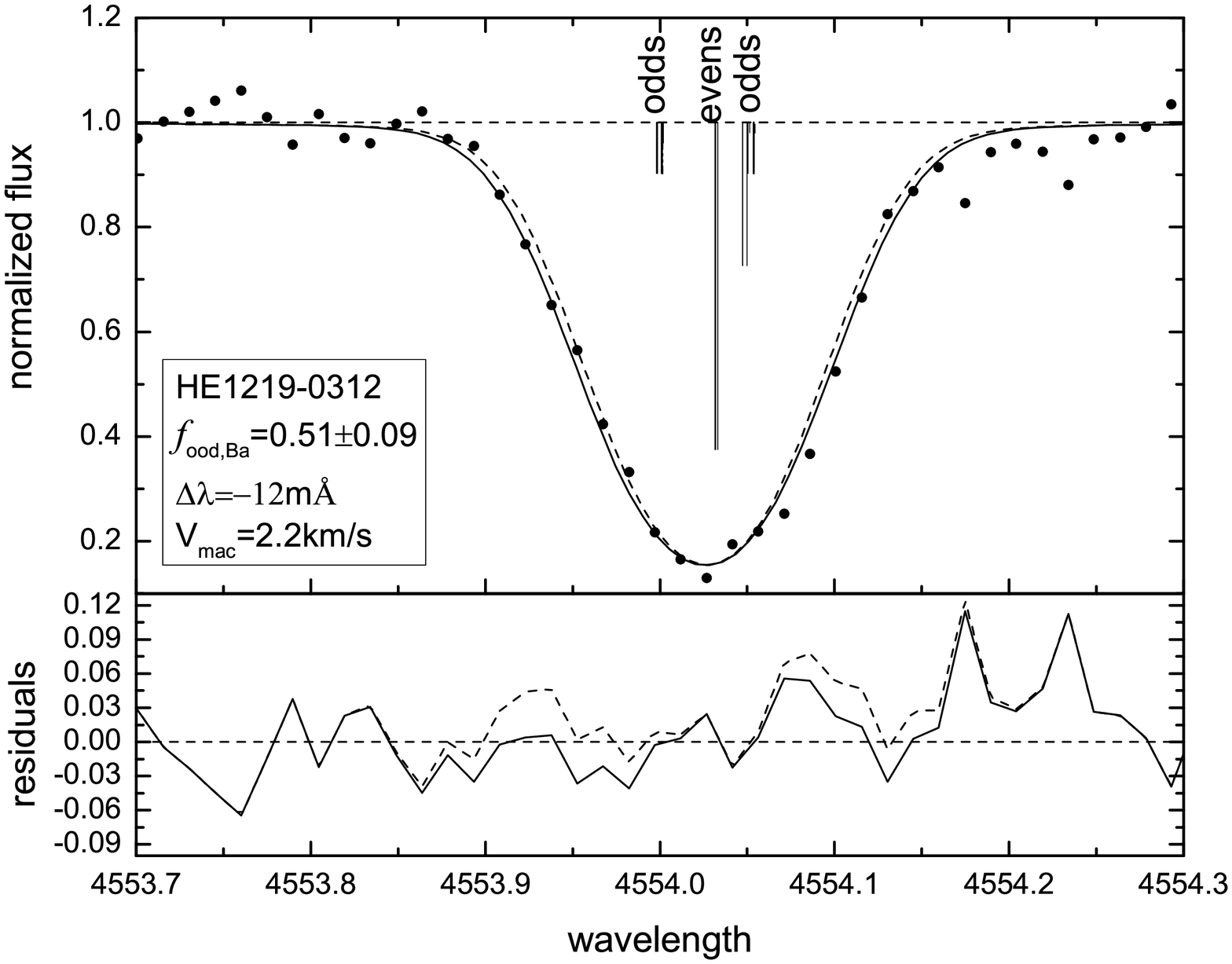}{0.4\textwidth}{}
 \hspace{-32mm}
          \fig{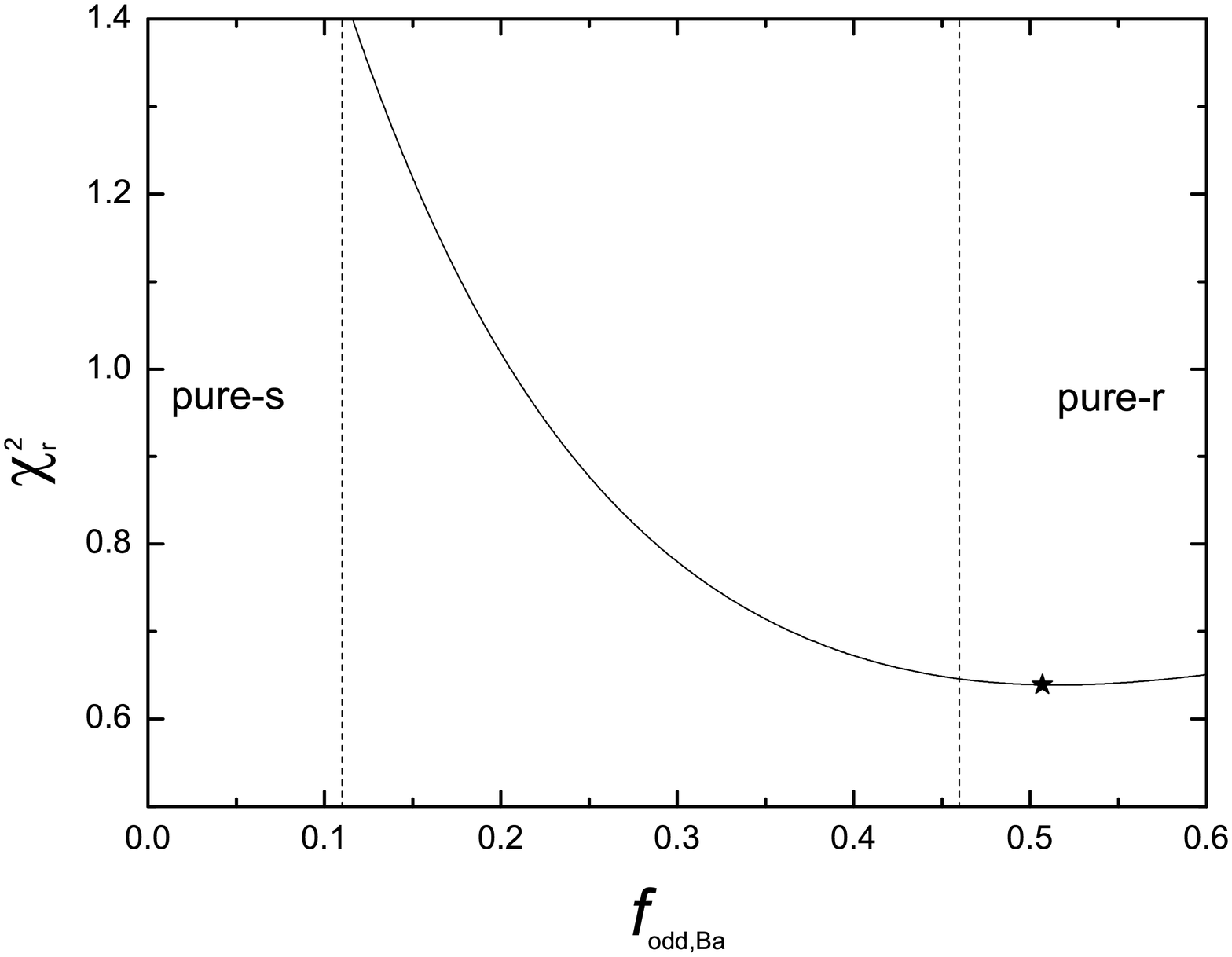}{0.4\textwidth}{}
      }    
           \vspace{-15mm}
\gridline{\fig{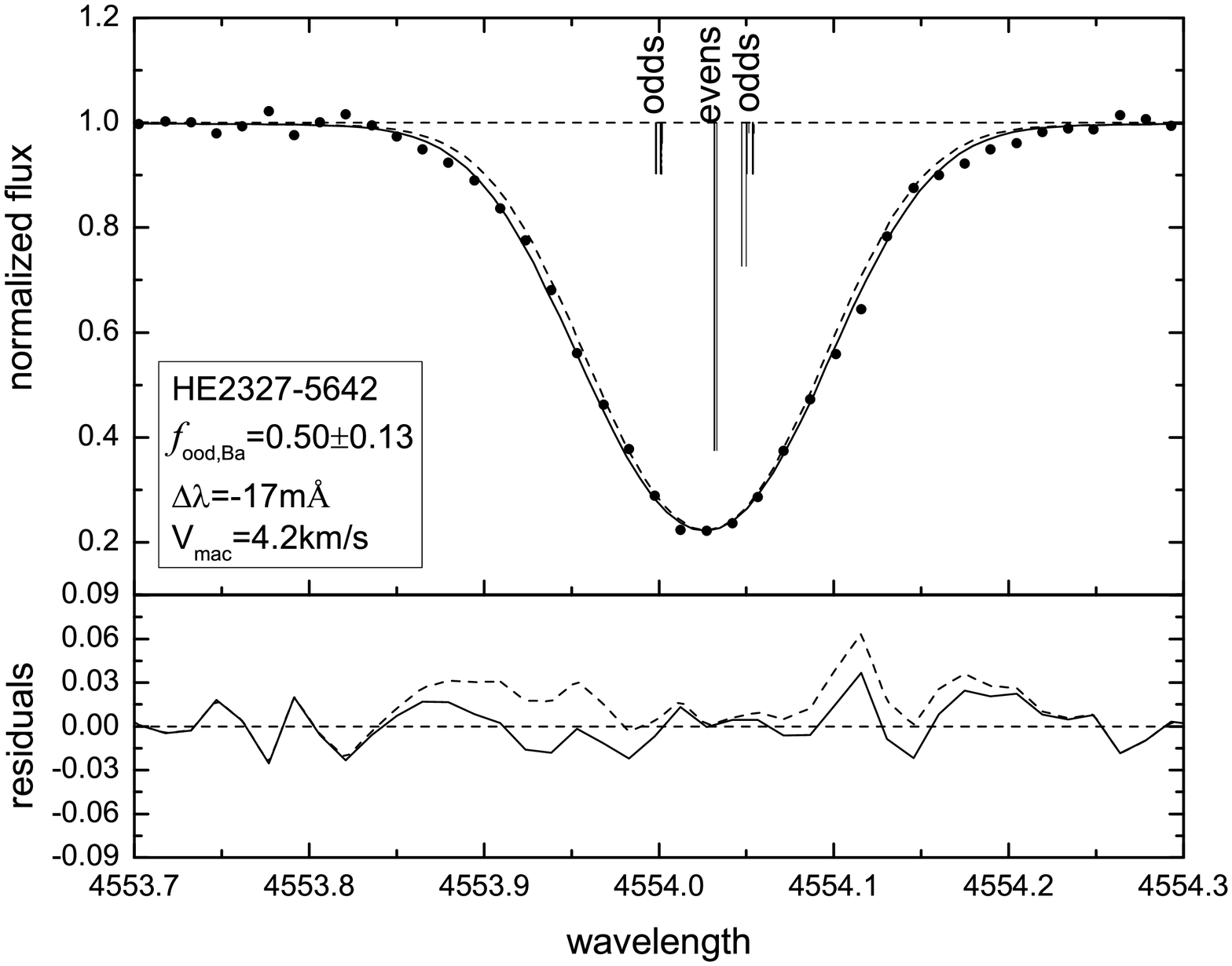}{0.4\textwidth}{}
 \hspace{-32mm}
          \fig{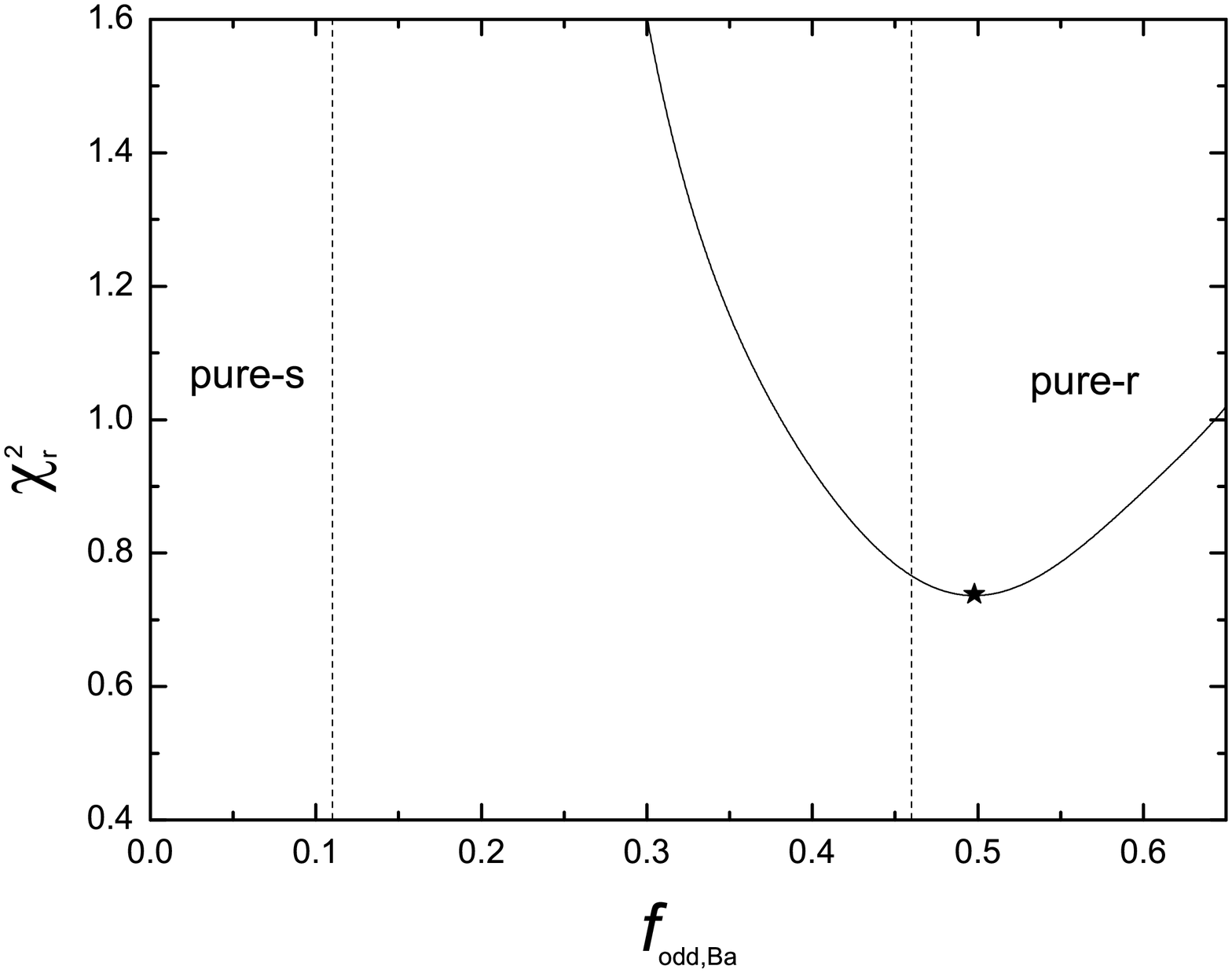}{0.4\textwidth}{}
      } 
           \vspace{-15mm}
\gridline{\fig{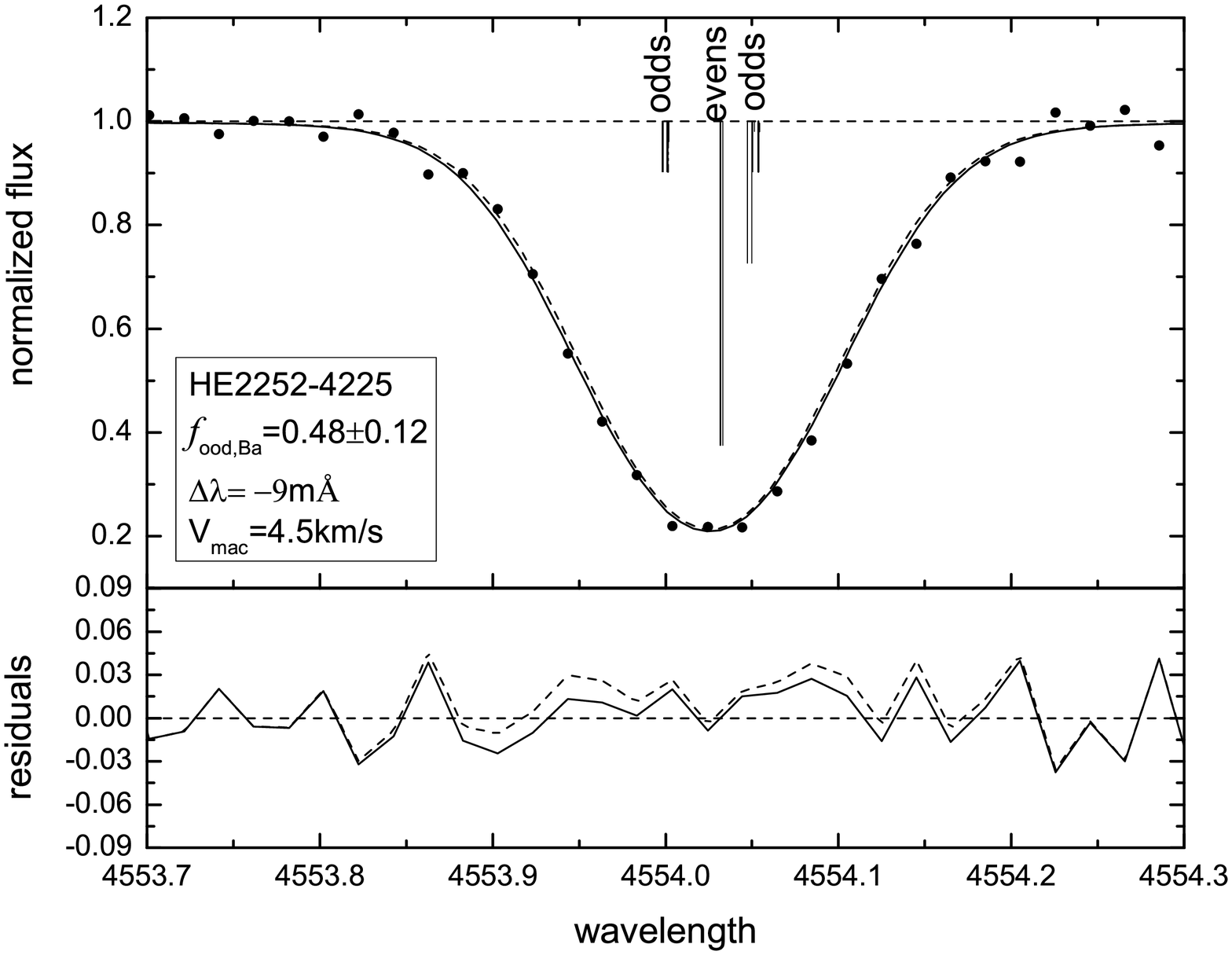}{0.4\textwidth}{}
 \hspace{-32mm}
          \fig{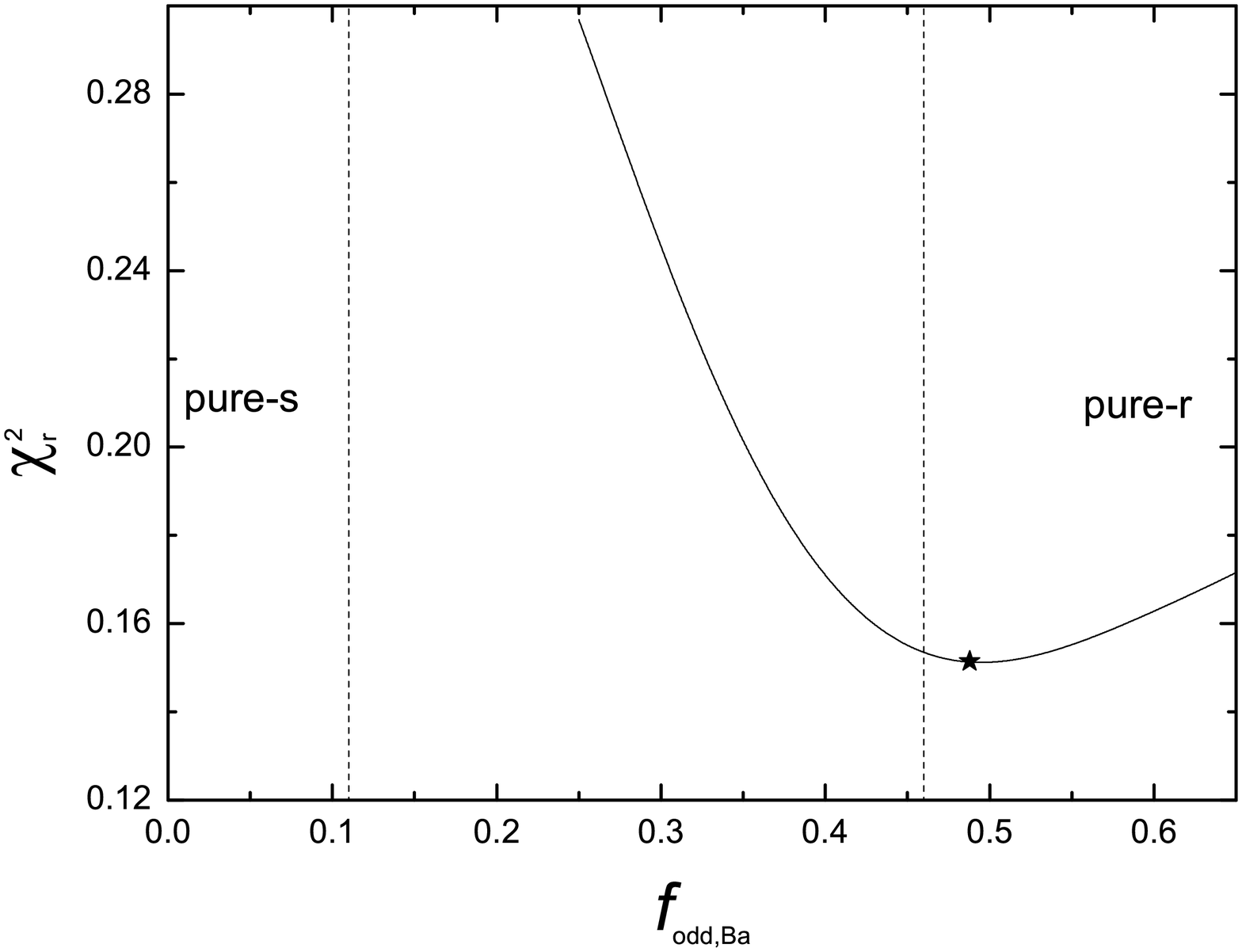}{0.4\textwidth}{}
      }   
           \vspace{-5mm}
\caption{Left panel (from up to down): the best statistical fit synthetic profile obtained with \fodd\,$=0.46, 0.51, 0.50, 0.48$ and 
NLTE line shapes for the observed (filled circles) Ba\,II resonance line of $\lambda4554$ in \cs, \hen, \hes\ and \het, respectively. 
 The residual plots are shown below of each subfigure, respectively. For comparison, the lines with \fodd\,$=0.30$ ($0.46-2\sigma$), 0.33 ($0.51-2\sigma$),
 0.37 ($0.50-\sigma$), 0.36 ($0.48-\sigma$) and the corresponding residuals
  have been plotted (dash-dot line) in each subfigures. The value for $V_\mathrm{mac}$ has been optimised to one that 
  minimises $\chi_\mathrm{r}^2$, and the value for [Ba/Fe] remains the same for each star. 
  Right panel (from up to down): we show the $\chi_\mathrm{r}^2$ fit for the $\lambda4554$ line, every star
  shows where the minimum of the fit lies. The vertical dotted lines indicate the solar $f^s_{\mathrm{odd,Ba}}$ (left)
  and $f^r_{\mathrm{odd,Ba}}$ (right) calculated from \citet{arl99}, respectively.} \label{fitting4554}
\end{figure*}

From Table~\ref{nltefodd},
we can see that for most of sample stars our NLTE Ba abundances are slightly higher than the LTE ones
from literatures except \het, which is lower than its LTE abundance about 0.13\,dex.
This can be naturally explained by the NLTE effects on different kind of Ba\,II lines.
\citet{mas14} used the subordinate Ba\,II lines of $\lambda5853$, $\lambda6141$ and 
$\lambda6496$ determining the LTE Ba abundance for \het. Although the subordinate 
lines are almost free of HFS effects, it is found that 
large negative NLTE abundance corrections for these lines, $\Delta\mathrm{NLTE}=-0.02$, 
$-0.17$ and $-0.23$\,dex, respectively. In fact, our NLTE Ba abundance $\mathrm{[Ba/Fe]}=0.16\pm0.08$
is consistent with the NLTE result $\mathrm{[Ba/Fe]}=0.14\pm0.04$ obtained by
\citet{mas14}. The NLTE effects on the resonance Ba\,II line of $\lambda4554$ is strong,
resulting a positive NLTE abundance correction, $\Delta\mathrm{NLTE}\approx+0.2$
\citep{asp05}.
%For \het, the large difference between our NLTE Ba abundance and the LTE
%one determined by \citet{mas14} is due to the adopting of Ba\,II subordinate lines,
%$\lambda5853$ and  $\lambda6496$ in this work. \citet{mas14} used the subordinate lines
%$\lambda5853$, $\lambda6141$ and $\lambda6496$ to determine the LTE Ba abundance.
%Although the subordinate lines are almost free of HFS effects, \citet{mas14} presented
%out large negative NLTE abundance corrections for these lines, $\Delta\mathrm{NLTE}=-0.02$, 
%$-0.17$ and $-0.23$\,dex, respectively. In fact, our NLTE Ba abundance $\mathrm{[Ba/Fe]}=0.16\pm0.08$
%is consistent with the NLTE result $\mathrm{[Ba/Fe]}=0.14\pm0.04$ determined by
%\citet{mas14}. Although the large difference between our NLTE Ba abundance and the LTE
%one determined by \citet{mas14} in \het, their NLTE abundane
%$0.14\pm0.04$ is consistent with our NLTE value $0.16\pm0.08$.
For \cs, \hen\ and \hes, both the resonance Ba\,II line of $\lambda4554$ and the subordinate lines
such as Ba\,II $\lambda5853$, $\lambda6496$ etc. were used to determine 
their LTE Ba abundances by \citet{hay09,mas10}. NLTE largely
reduced the difference in LTE abundances between different Ba\,II lines, resulting the slightly
higher Ba abundances for \cs, \hen\ and \hes.

\subsection{Ba Odd-Isotope Fractions\label{frac}}

The \fodd\ values for \cs, \hen, \hes, and \het\ were derived through fitting the profiles of Ba\,II resonance line
$\lambda4554$, and listed in Table~\ref{nltefodd}. 
The Ba atomic model from \citet{mas99} were adopted, and
the hyperfine structure components of Ba\,II 4554 can be found in \citet{mas06}.
Figure\,\ref{fitting4554} shows the determining processes for these stars, 
and our calculations of the synthetic profile are based on the NLTE line formations for the 
Ba\,II resonance line of $\lambda4554$.
\begin{table*}[!hbtp]
\begin{center}
\caption{Effects on the values of \fodd\ resulting from uncertainties of atomic data and stellar parameters.}
\label{uncertainty}
\begin{tabular}{lcccccccc}
\hline
\hline
Input & Input &CS\,29491 &Input& HE\,1219&Input& HE\,2327&Input& HE\,2252 \\
parameter&error &-069 &error&-0312  &error&-5642 &error&-4225\\
\hline
log\,$C_6(5d-6p)$ &	+0.1 &	-0.01 &	+0.1	&	+0.04	&	+0.1	&	-0.08 &	+0.1	&	-0.07	\\   
$\mathrm{[Fe/H]}$ &	-0.08 &	+0.03 &	-0.08	&	+0.06	&	-0.08	&	+0.06 &	-0.08	&	+0.01	\\   
$T_\mathrm{eff} $ &	  +80 &	+0.05 &	  +80	&	+0.02	&	  +80	&	+0.04 &	  +80	&	+0.02	\\   
log\,$g$                &	-0.20 &	-0.01 &	-0.20	&	-0.02	&	-0.15	&	-0.04 &	-0.15	&	+0.01	\\   
$V_\mathrm{mic}$  &	-0.10 &	+0.05 &	-0.10	&	-0.01	&	-0.10	&	+0.01 &	-0.10	&	-0.01	\\   
log$gf$(4554)     &	-0.1 &	+0.02 &	-0.1 &	+0.04	&	-0.1 &	+0.04 &	-0.1 &	+0.06	\\   
log\,$C_6$(4554)  &	+0.1 &	+0.01 &	+0.1 &	-0.01	&	+0.1 &	+0.04 &	+0.1 &	-0.07	\\   
$V_\mathrm{mac}$  &	-0.2 &	+0.01 &	-0.2 &	-0.01	&	-0.2 &	+0.01 &	-0.2 &	+0.01	\\   
$\Delta$(total) & & $\pm0.08$ & & $\pm0.09$ & & $\pm0.13$ & & $\pm0.12$ \\
\hline
\vspace{-10mm}
\end{tabular}
\end{center}
\end{table*}

Following \citet{men16}, the macroturbulence values were still allowed to be free during 
the analysis processes. Our synthetic spectra were convolved with the profile broadening by 
macroturbulence $V_\mathrm{mac}$, the instrumental profile $\Gamma$ and the rotational velocity 
$v\,\mathrm{sin}\,i$. The projected rotational velocity $v\,\mathrm{sin}\,i=1.5$\,km\,s$^{-1}$
was adopted for all sample stars. Through a Gaussian fit to the Th-Ar
lamp spectrum,
% observed with the same instrumentation setup when the object exposures, 
we obtained the instrumental broadening values,
$\Gamma=3.1, 3.0, 3.6$\,km\,s$^{-1}$ for \cs, \hen, \het\ except \hes, as no proper 
lamp spectrum found for \hes. Then the macroturbulence values of the Ba\,II line at 4554\,\AA\ 
were found to be $V_\mathrm{mac} = 2.5, 2.2, 4.5$\,km\,s$^{-1}$ for \cs, \hen, and \het, 
respectively. As these three parameters, $V_\mathrm{mac}$, $\Gamma$ and
$v\,\mathrm{sin}\,i$ are convolved together when calculating the synthetic spectra
using the SIU software, we can not determine the $V_\mathrm{mac}$ value
for \hes\ due to no proper lamp spectra obtained. For \hes, we obtain a Gaussian value 
4.2\,km\,$^{-1}$(i.e., $\sqrt{V_\mathrm{mac}^2+\Gamma^2)}$). Finally, the $\chi_\mathrm{r}^2$ fits
also require small wavelength shift, $\Delta\lambda=-8, 12, -17$, and $-9$\,m\AA, respectively.
Figure~\ref{fitting4554} shows the best statistical fit to the $\lambda4554$ line and residual 
(synthetic $-$ observed profile) in the left panel from up to down for \cs, \hen,
\hes, and \het, respectively. The synthetic profiles for Ba\,II line of $\lambda4554$ with 2$\sigma$
for \cs\ and \hen, or 1$\sigma$ for \hes\ and \het\ lower than their final \fodd\ value 
were also been presented for comparison. From the residual plots 
for the Ba\,II line of $\lambda4554$, we can see that the fits of the synthetic lines 
with 2$\sigma$ or 1$\sigma$ deviation from their final \fodd\ values are poor.

In the right panel of Figure~\ref{fitting4554}, 
we show the $\chi_\mathrm{r}^2$ versus \fodd. 
The $\chi_\mathrm{r}^2$ minimum 0.272, 0.639, 0.736, and 0.151 can be respectively obtained at 
\fodd\, $=0.46, 0.51, 0.50, 0.48$ for \cs, \hen, \hes, and \het, where the gradient of the 
$\chi_\mathrm{r}^2$ curve is zero. We note that all of the four r-II stars are
enhancement with Ba, [Ba/Fe]\,$=0.26, 0.72, 0.34, 0.16$, and their Ba\,II resonance
line of $\lambda4554$ are strong enough, $EW=124.1, 136.9, 127.2, 151.8$\,m\AA\ 
(see Table~\ref{ba.abundance}), to get reliable \fodd\ values.

%\begin{figure*}
%\plottwo{cs.pdf}{cskai.eps}
%\caption{Left panel: the best statistical fit synthetic profile obtained with \fodd\,$=0.23$ and 
%NLTE line shapes for the observed (filled circles) Ba\,II resonance line at 4554\,\AA\ in \cs\ 
%  with the residual plots below. For comparison, a line with \fodd\,$=0.11$ (i.e. $0.23-\sigma$) and residual
%  have been plotted (dash-dot line). The value for $V_\mathrm{mac}$ has been optimised to one that 
%  minimises $\chi_\mathrm{r}^2$, and the value for [Ba/Fe] remains the same. 
%  Right panel: we show the $\chi_\mathrm{r}^2$ fit for the 4554\,\AA\ line, the star
%  shows where the minimum of the fit lies.\label{fitting4554}.}
%\end{figure*}

\subsection{Uncertainty of the Ba-odd-isotope Fraction}

Random errors in \fodd\ are resulted from the errors in Ba abundance and the stellar parameters, i.e., $T_\mathrm{eff}$, log\,$g$ 
and $V_\mathrm{mic}$, while the uncertainty of the Ba\,II $\lambda4554$ atomic parameters, i.e., log\,$gf$ and log\,$C_6$, causing 
their systematical errors. Following \citet{men16}, we estimated the total errors of \fodd\ values as 
$\pm0.08$ for \cs, $\pm0.09$ for \hen, $\pm0.13$ for \hes, and $\pm0.12$ 
for \het, respectively. Here, various sources of uncertainties influencing the derived 
\fodd\ values were included, which were listed in Table~\ref{uncertainty}.
Although we can not determine the $V_\mathrm{mac}$ value
for \hes\ without proper lamp spectra obtained, a same uncertainty value $-0.2$ of the 
$V_\mathrm{mac}$ was adopted to compare with other sample stars
when estimating its total errors in \fodd.

\begin{figure*}
\gridline{\fig{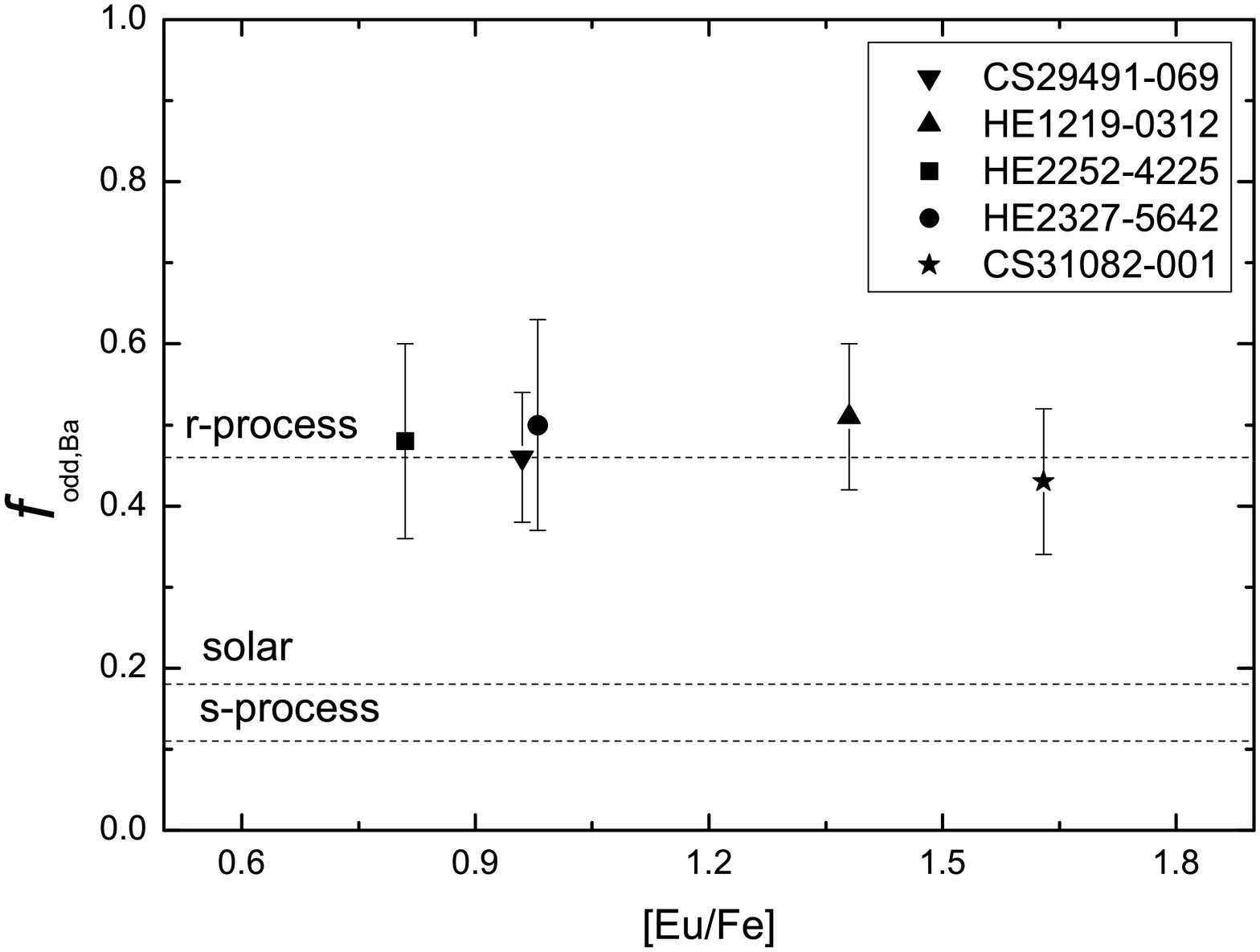}{0.4\textwidth}{a}
 \hspace{-32mm}
          \fig{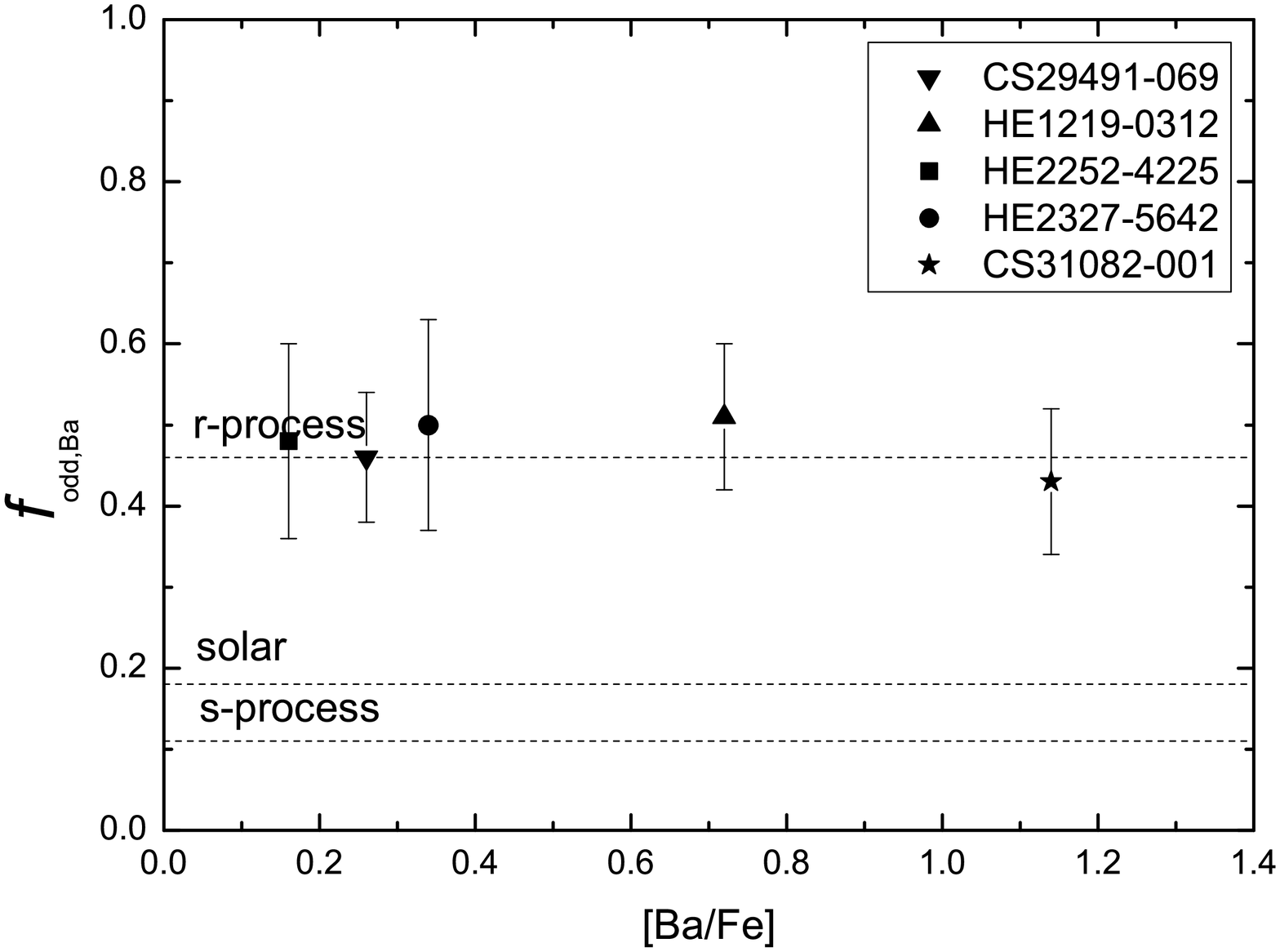}{0.4\textwidth}{b}
          }
           \vspace{-6mm}
\gridline{\fig{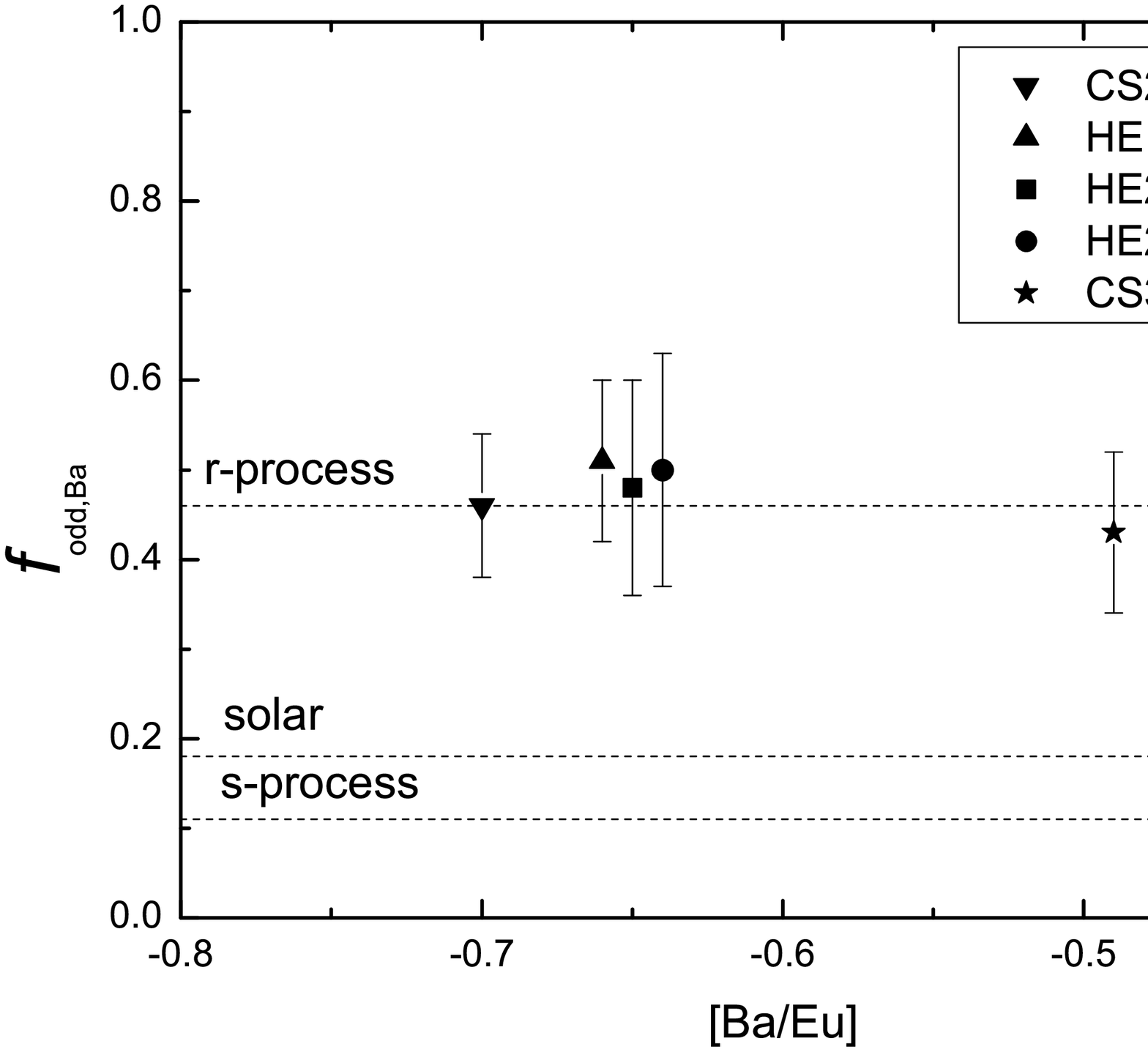}{0.4\textwidth}{c}
 \hspace{-32mm}
          \fig{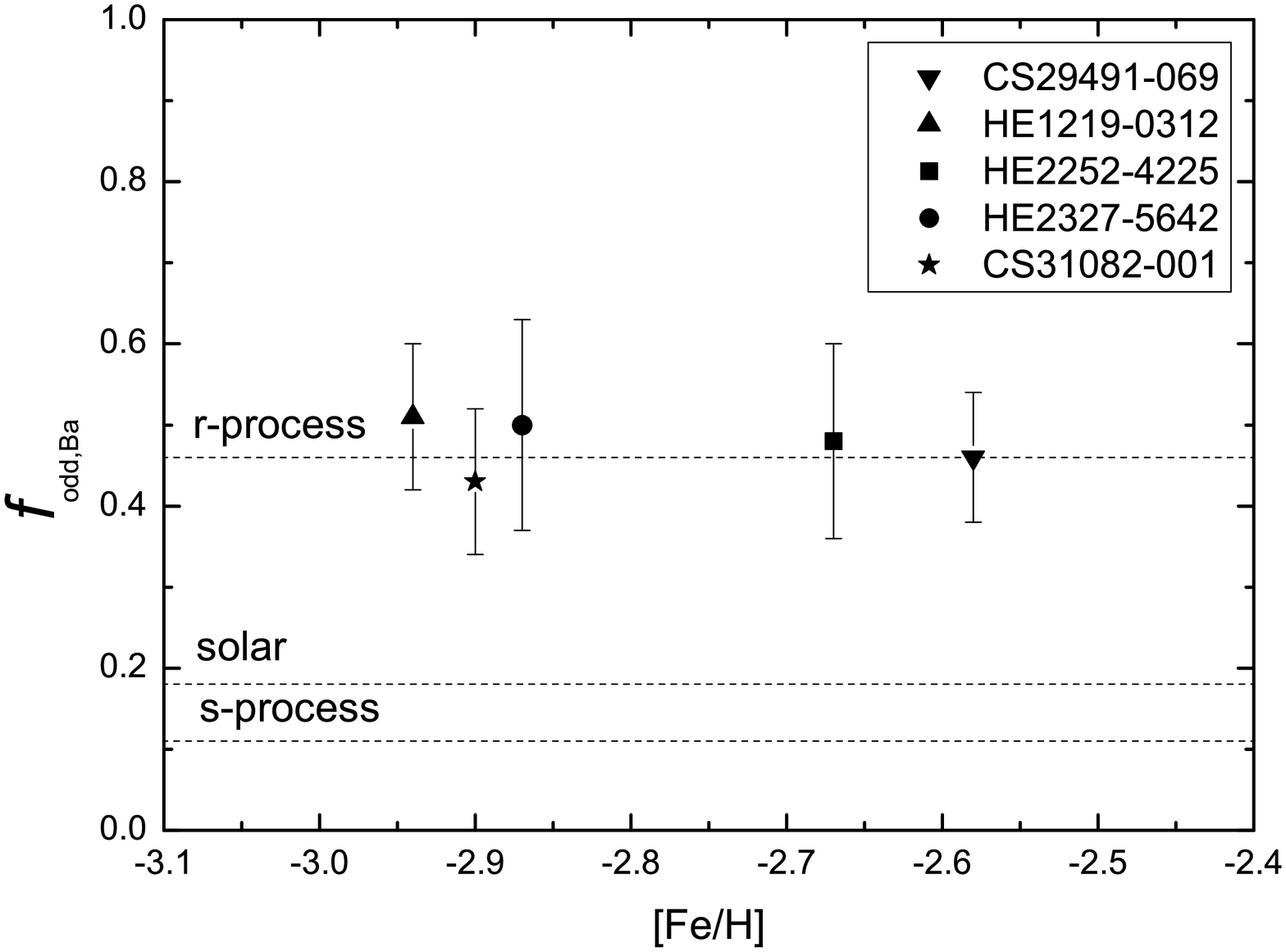}{0.4\textwidth}{d}
      }    
           \vspace{-2mm}
\caption{The fraction of the odd Ba isotopes, \fodd, versus a: [Eu/Fe] (LTE), b: [Ba/Fe] (NLTE), c: [Ba/Eu], and 
d: [Fe/H] (NLTE except CS\,31082-001). Filled upsidedown triangle represents \cs, filled triangle for \hen,  
filled square for \het, and filled circle for \hes, respectively.
The \fodd\ value of CS\,31082-001 (asterisk) derived by \citet{men16} was also presented out for comparing.
Uncertainties are shown by short vertical lines. Dotted horizontal lines indicate the \fodd\ values, 
0.18 for the solar system, 0.46 for the pure r-process, and 0.11 for the pure s-process in the 
solar barium abundance predicted by \citet{arl99}. The Eu abundance of \cs\ and \hen\ adopted from
\citet{hay09}, from \citet{mas10} for \hes, and from \citet{mas14} for \het.
\label{fbaeu}.}
\end{figure*}

\section{Origin of Heavy Elements for r-II stars}

The fraction of the odd Ba isotopes, \fodd, in our sample stars 
versus a: [Eu/Fe], b: [Ba/Fe], c: [Ba/Eu], and d: [Fe/H] are shown in Figure~\ref{fbaeu}.
As comparison, the \fodd\ value of CS\,31082-001 derived by \citet{men16} was also included.
In Figure~\ref{fbaeu}a, the adopted LTE values of [Eu/Fe] are $0.96\pm0.10$ for \cs\, \citep{hay09}, $1.38\pm0.10$
for \hen\, \citep{hay09}, $0.98\pm0.11$ for \hes\, \citep{mas10}, and $0.81\pm0.08$ for \het\, \citep{mas14}.
In addition, \citet{mas14} also provided a NLTE [Eu/Fe] $0.91\pm0.09$ for \het.

It is noted that the adopted {[Eu/Fe] values for \cs, \hes, and \het\ are all slightly lower than 1.0 \citep[one 
of the two criteria suggested for r-II stars by ][]{bee05}. \citet{mas10} found 
a clear distinction in abundance ratios [Sr/Eu] between the r-II and r-I \citep[$0.3\le\mathrm{[Eu/Fe]} <1.0$ 
and $\mathrm{[Ba/Eu]}<0$, see][]{bee05} stars. The r-II stars show a low [Sr/Eu] value,
that is, $\sim-0.92\pm0.13$, however, the r-I stars have 0.36\,dex higher [Sr/Eu] values,
i.e., $\sim-0.56\pm0.13$. Therefore, \citet{mas10} proposed to use $\mathrm{[Sr/Eu]}<-0.8$ as the third criterion to 
identify the r-II stars.
%, that is, $\mathrm{[Sr/Eu]}<-0.8$, in addition to the two, $\mathrm{[Eu/Fe]}>1.0$ and $\mathrm{[Ba/Eu]}<0$,
%as suggested by \citet{bee05}. %Considering the [Sr/Eu] values, the above three sample stars
%with slightly lower [Eu/Fe] should belong to the r-II group.
The LTE values of [Sr/Eu] for our sample stars were calculated
from literatures, which are $-0.81\pm0.21$ for \cs, $-1.03\pm0.21$ for 
\hen\, \citep{hay09}, $-1.13\pm0.14$ for \hes\, \citep{mas10}, $-0.94\pm0.02$ 
(LTE) and $-1.0\pm0.06$ (NLTE) for \het\, \citep{mas14}. 
%Based on the [Sr/Eu] values of 
%\cs, \hes, and \het, 
Thus, these stars should be divided into the r-II group. 
\hen\, and CS\,31082-001 ($\mathrm{Eu/Fe]}= 1.63\pm0.11$) are usually regarded as the benchmark r-II stars \citep{mas14}.

The r-II stars have very similar abundance patterns for the elements in the range from Ba to Pb, thus 
a common origin in the r-process for the heavy elements ($56\le Z \le82$) 
has been suggested for these stars \citep[see][]{sne08,cow11,mas14}.
%Figure~\ref{fbaeu} shows the fraction of the odd Ba isotopes, \fodd, in our sample stars 
%versus (a) [Eu/Fe], (b) [Ba/Fe], (c) [Ba/Eu], and (d) [Fe/H].
%For comparing, the \fodd\ value of CS\,31082-001 derived by \citet{men16} was also presented out.
From Figure~\ref{fbaeu}, we can see that all of the \fodd\ values,
$0.46\pm0.08$, $0.51\pm0.09$, $0.50\pm0.13$, $0.48\pm0.12$, $0.43\pm0.09$ for \cs, \hen, \hes, \het, 
and CS\,31082-001, are consistent with
the value $0.46$ of \foddr\ well within the error bars,
%, which are also independent of their
%abundance ratios [Eu/Fe], [Ba/Fe], [Ba/Eu], and [Fe/H]. 
which supports the ``stellar model" predicted result of \citet{arl99}.
Using the formula, 
r-process(\%) $= (f_\mathrm{odd,Ba}-0.11)/0.0035$ \citep[see][]{gal10,men16}, the r-process contributions
can be calculated as $100\pm22.9\%$ for \cs, $114.3\pm25.7\%$ for \hen, $111.4\pm37.1\%$
for \hes, and $91.4\pm25.7\%$ for \het, respectively. As the value of the r-process 
contribution larger than $100\%$ is not physical, we adopted $100\%$ for both \hen\ and \hes.
The r-process contribution is $91.4\pm25.7$\% for CS\,31082-001 adopted from \citet{men16}.
This indicates that the r-process produced almost all of the Ba element in these r-II stars.
It is noted that the Ba element usually is regarded as the representative element of s-process, 
thus we can infer that almost all of the heavy elements (at least beyond Ba and up to Pb) 
in these r-II stars are synthesized by a single nucleosynthesis process, that is, the main r-process.
%which should also be the same in any r-II stars. 
%The range of the heavy elements referred above
%is deduced from the observation results that the r-II stars have very similar abundance patterns 
%of the elements in the range from Ba to Pb. Hence, neutron-capture elements beyond Sr and up to Ir (77) in HE 2252?4225 have a common origin in the classi- cal main r-process. 
Based on the fact that all of \fodd\ values for our five r-II sample stars are close to the 
solar pure r-process value 0.46, we can speculate that this may be also same for other
r-II stars. It can be seen that the conclusion for the origin of neutron-capture elements
obtained from their abundance pattern is also supported on the isotopic level.

\section{Characters of \fodd\ in r-II Stars}

From Figure~\ref{fbaeu}, we can see that the \fodd\ values of the r-II stars are very good agreement with 
that of the solar pure r-process \foddr, although they suffer large scatters, about 0.9\,dex and 1.0\,dex, in the abundance ratios 
[Eu/Fe] (Figure~\ref{fbaeu}a) and [Ba/Fe] (Figure~\ref{fbaeu}b), respectively. This means that
the main r-process responsible for the Eu and Ba abundance of the r-II stars has a intrinsic \fodd\ value, about 0.46,
which is very close to the mean value about $0.48\pm0.01$ of \fodd\ in our r-II sample stars,
and the environment for the main r-process is quite robust.
In other words, the \foddr\ value obtained from the solar abundance by residul method is 
reliable. This also indicates that the different enhancement level of Eu and Ba, including other
relative heavy elements in the r-II stars may be due to the dilution effects.% by the clouds before they formed. 
%Figure~\ref{fbaeu}a,b show the r-II sample stars have large scatter, about 0.9\,dex and 1.0\,dex, in the abundance ratios 
%[Eu/Fe] and [Ba/Fe], respectively.

Figure~\ref{fbaeu}c shows that \fodd\ versus [Ba/Eu] for r-II stars. 
From this figure, we can see that these r-II stars have a small scatter about 0.08\,dex 
among \fodd\ values, and the scatter of about 0.20\,dex among [Ba/Eu] ratios, which 
even reach to about 0.30\,dex in nine r-II stars collected by \citet{mas10}. 
Furthermore, the mean \fodd\ value is $0.48\pm0.01$, which is almost equal to
the solar pure r-process value 0.46. 
%As 82\% of the solar Ba is produced by the s-process and 94\% of the solar Eu is produced
%by the r-process, Eu and Ba are usually regarded as the representative elements of the 
%r-process and s-process, respectively. Therefore, the abundance ratio [Ba/Eu] is an 
As Ba and Eu are the representative elements for the s- and r-process, respectively, 
the abundance ratio [Ba/Eu] is regarded as an 
important indicator % during the studies of the abundance analysis, which can help us 
to understand the relative importance 
of the r- and s-process, throughout the Galaxy history or in 
a star with peculiar abundances of neutron-capture elements well. For r-II stars, 
\citet{mas10} gave a mean value of $\mathrm{log(Ba/Eu)} = 1.05\pm 0.10$ calculated from nine r-II stars, 
which corresponds $\mathrm{[Ba/Eu]}=-0.61\pm0.10$. This value is about 0.1\,dex larger than the solar pure r-process 
value $\mathrm{log(Ba/Eu)} =0.93$, which corresponds $\mathrm{[Ba/Eu]_r}=-0.73$ \citep{arl99}.
This can be explained as that a small number of s-nuclei existed in the matter out of which the r-II 
stars formed \citep{mas10}. For r-II stars, the mean \fodd\ seems to show a better agreement with that of the solar
pure r-process value \foddr, while the difference is slightly large between the mean abundance ratio %than the mean abundance ratio
[Ba/Eu] and the solar 
pure r-process value $\mathrm{[Ba/Eu]_r}$. This indicates that \fodd\ is also an important indicator
to study the relative contributions of the r- and s-process, 
%which is not only because 
%the better agreement between the mean \fodd\ and  \foddr, but also because
%nucleosynthesis information provided by \fodd\ in more detail on the isotopic level than [Ba/Eu] 
%on the abundance level.
which can provide more detailed nucleosynthesis information on the isotopic level than [Ba/Eu]
on the abundance level.

In addition, the r-contribution is $105.7\pm2.9\%$ calculated from the mean \fodd\
$0.48\pm0.01$ using the formula referred in Section 5.%, r-process(\%) $= (f_\mathrm{odd,Ba}-0.11)/0.0035$.
The physical value 100\% for the r-contribution to r-II stars should be adopted,
which means that almost all of their heavy elements from Ba to Pb
were produced by the r-process.
%Figure~\ref{fbaeu}d shows that \fodd\ versus [Fe/H] for r-II stars. From Figure~\ref{fbaeu}d,
%we can see that there is no trend of \fodd\ with the metallicity increasing. 
We plot \fodd\ versus [Fe/H] for r-II stars in Figure~\ref{fbaeu}d, and found
no trend of \fodd\ with the metallicity.
This also supports that
the \fodd\ value, i.e. 0.46, calculated from the solar r-residul is the typical value with great possibility
for the r-process responsible for the production of heavy elements beyond Ba and up to Pb in r-II stars,
and it is universal and stable throughout the whole Galaxy history.

\section{Conclusions}

In this work, we determined the \fodd\ values, $0.46\pm0.08$, $0.51\pm0.09$, 
$0.50\pm0.13$, $0.48\pm0.12$ for four r-II stars, \cs, \hen, \hes\, and \het, based on the 
high resolution and high signal-to-noise spectra, which are downloaded from ESO archive.
Using the formula, r-process(\%) $= (f_\mathrm{odd,Ba}-0.11)/0.0035$, the r-contributions
to these r-II stars are calculated as $100\pm22.9\%$ for \cs, $114.3\pm25.7\%$ 
for \hen, $111.4\pm37.1\%$ for \hes, and $91.4\pm25.7\%$ for \het, respectively.
As the value of the r-contributions larger than $100\%$ is not physical, we adopted 
$100\%$ for both \hen\ and \hes. We confirmed that almost all of the 
heavy elements (from Ba to Pb) in r-II stars have a common origin, that is, from
a single r-process (the main r-process), and the r-process environment should be quite
robust. 

We found that the different enhancement level of Eu and Ba, including other
relative heavy elements in the r-II stars should be mainly due to the dilution effects 
by the clouds before they formed, as their \fodd\ shows a intrinsic nature and has 
no trends with [Eu/Fe] and [Ba/Fe] ratios. 
In addition, we also found that the \fodd\ in r-II stars has no trends with the metallicity.
Thus, we inferred that the \foddr\ value for the main r-process is stable and universal 
throughout the whole Galaxy history, which is about 0.46.
% calculated from 
%the solar r-residual.

Our results suggest, except [Ba/Eu], \fodd\ is also an important indicator to study the relative importance
of the r- and s-process during the chemical evolution history of the Milky Way or
the enhancement mechanism of the abundance peculiar stars for neutron-capture 
elements. We found that comparing to [Ba/Eu], \fodd\ in r-II stars has smaller 
scatter about 0.08\,dex with a mean \fodd\ of
$0.48\pm0.01$, shows good agreement with the solar pure r-process value 0.46. 
In addition, the mean \fodd\ in r-II stars supports that 
the r-process contributed 100\% of their heavy elements beyond Ba and up
to Pb, which is slightly different with that only small contribution from the s-process
for the s-nuclei deduced from their mean [Ba/Eu] \citep{mas10}. 
%Thus, 
%we suggested that the \fodd\ should be a better indicator
%than [Ba/Eu] to study the enhancement history of neutron-
%capture elements in our Galaxy or a star.

The isotopic fractions including \fodd, $f_{151}$, etc. are very important 
observational constrains for the theoretical calculations of the main r-process,
which provides the opportunity to further and reliably identify the origin site from
type II supernova or neutron star merger events. It would be very interesting to 
perform more such kind measurements in future to shed light on this issue.
%such kind of calculations under the isotopic constrains in the future in order to shed light on this issue.}

%This agreement suggests that a similar process is responsible for generating the heavy r-elements, both for the most metal- poor stars and for the Sun.
%Intriguingly, the robust nature of the abundance pattern in these halo stars, born perhaps before the Galaxy was fully formed, suggests a possibly more widespread agreement or relative consistency among our Galaxy and others.

\acknowledgments

We thank the anonymous referee for positive and constructive comments that greatly 
helped to improve this paper. W.Y. Cui acknowledges Dr. H.L. Yan for the help on spectra reducing. 
This work is supported by  the National Natural 
Science Foundation of China under grants 11643007, 11773009, 11547041, 11673007, 11473033.

\software{SIU (Reetz 1991)}

\end{document}